\documentclass[nonamebreak,options]{jaa}

\newcommand{\arcmin}{\ensuremath{^{\prime}}}
\newcommand{\arcsec}{\ensuremath{^{\prime\prime}}}
\usepackage[utf8]{inputenc}
\usepackage{aas_macros}
\usepackage{xcolor}
\usepackage{natbib}
\usepackage{amsmath}
\usepackage{graphicx}
\usepackage{authblk}
\usepackage{hyperref}
\usepackage{placeins}    
\usepackage{lscape}  
\usepackage{booktabs}
\usepackage{textcomp} 
\usepackage{array}
\usepackage{tabularx}
\usepackage{txfonts}
\usepackage{lipsum}
\usepackage{amsmath}
\usepackage{subcaption}  

\begin{document}\sloppy

\title{Stellar contents and Star Formation in IRAS 18456-0223}


\author{Nilesh Pandey\textsuperscript{1,2,3,*} and U. S. Kamath\textsuperscript{4}}

\affilOne{\textsuperscript{1}Indian Institute of Science Education and Research Thiruvananthapuram, Maruthamala P.O., Vithura 695551, India.\\
\textsuperscript{2}Astronomical Institute of the Academy of Sciences of the Czech Republic, Fričova 298, CZ-251~65 Ondřejov, Czech Republic. \\
\textsuperscript{3}Astronomical Institute, Charles University in Prague, V Holešovičkách 2, CZ-180~00 Praha~8, Czech Republic.\\
}
\affilTwo{\textsuperscript{4}Indian Institute of Astrophysics, Koramangala II Block, Bengaluru 560034, India.}


\twocolumn[{

\maketitle

\corres{pandeynilesh51@gmail.com}

\msinfo{11 August 2025}{16 February 2026}


\begin{abstract}
We use various analytical techniques to study Young Stellar Objects (YSOs) in an area of approximately \(10~\arcmin \times 10~\arcmin\) in the IRAS 18456-0223 star-forming region. We use archival optical (\textit{Gaia DR3}) and infrared (\textit{2MASS}, \textit{UKIDSS}, \textit{Spitzer}, \textit{WISE}, and \textit{Herschel}) data and our optical spectroscopy of three bright stars for this purpose. We identify 89 YSOs (80 class II and 9  class I), based on their infrared properties. Our multiwavelength SED fits of select YSOs show that they have masses $\sim0.1$–$7.2$~$M_\odot$ and are upto $4$~Myr old. Our Minimum Spanning Tree (MST) analysis shows that these YSOs, situated at around 600 pc, form clusters with radial extent of order $0.5\,\mathrm{pc}$ and mean surface densities $\sim60\,\mathrm{pc}^{-2}$. We compare \textit{UKIDSS} and \textit{2MASS} data of the YSOs and find that some of them show variability.
We constructed maps based on \textit{Herschel} data which reveal multiple column density peaks ($N_{\rm H_2}\sim 10^{22}\,\mathrm{cm}^{-2}$) embedded in cold ($T_d \sim 10{-}13$\,K) filaments. Our near-infrared extinction map exhibits several high-$A_V$ peaks, some of which coincide with the sub-mm column density maxima. Using our optical spectra of three bright sources, we show that they are of A-K type. One star shows the  Li I 6707\,\AA\, line, indicating its youth.
\end{abstract}

\keywords{stars: formation -- stars: protostars -- ISM: dust, extinction -- infrared : stars}

}]


\doinum{12.3456/s78910-011-012-3}
\artcitid{\#\#\#\#}
\volnum{000}
\year{0000}
\pgrange{1--}
\setcounter{page}{1}
\lp{1}

\section{INTRODUCTION}
\label{sec:intro}
Dust and gas in the environment surrounding young stars results in a large extinction along the line of sight, precluding fruitful observations in the optical. These regions are better studied at infrared wavelengths. One way of identifying potential star-forming regions is by making use of IRAS data. Low- and intermediate mass T Tau stars as well as the massive OB stars and their resulting H II regions can be identified based on their position in the IRAS two-colour diagram. (\citet{1988MNRAS.235..441H} ; \citet{1989ApJ...340..265W}). Observations of such IRAS-selected sources chiefly in the infrared have uncovered clusters of young stars at different scales and advanced our understanding about the structure and dynamics of star forming regions \citep{1993ApJ...407..657C}.

Some young stars called FUors (named after the prototype FU Orionis) show 5-6 magnitude outbursts in the optical that last for decades \citep{1977ApJ...217..693H}. Another class somewhat inaccurately termed as EXors (after the protoptye EX Lupi) exhibit smaller and shorter outbursts \citep{2008AJ....135..637H} According to the most commonly accepted model for FUor and EXor eruptions, the brightening is due to enhanced accretion or transfer of matter from the circumstellar disk onto the stellar photosphere (for example, see the reviews by \citet{1996ARA&A..34..207H} ; \citet{2014prpl.conf..387A}). This accretion is episodic, the outbursts are more frequent and energetic during the early phases, and their frequency and strength decreases as the star evolves towards the main sequence. Indeed, the nature of the circumstellar disc and the evolutionary state of a YSO determine its location on the $JHK$ two-colour diagram  \citep{1992ApJ...393..278L}. Thus FUors and EXors are secondary indicators of recent star formation. Although various star forming regions have been studied in detail only a dozen FUors were known a decade ago \citep{2015ApJ...801L...5K}. However, with the advent of many transient surveys, this number has shot up meaningfully. While the post-outburst phases are well-studied, the stellar and accretion properties of the progenitors of FUors are not well constrained. Therefore, it is not yet known whether the star returns to its pre-outburst state long after outburst.

\cite{2001A&A...370..991K} reported the discovery of a conspicuous flaring star coincident with IRAS 18456-0223 on a POSS I plate.  In the optical data, a very faint nebula and a compact group of stars in its vicinity could be seen. Using the \textit{DENIS} $K$ vs. $J$-$K$ diagram, they identified 28 YSO and T Tauri candidates in the region. The flaring star itself was not detected by \textit{DENIS}, leading them to assume that it appears enshrouded in a dust cloud with $A_V > 10$\,mag. Their distance estimate for this newly-discovered star-forming region was 1600$\pm$400 pc. This interesting region was not studied exclusively after that; the equatorial co-ordinates overlap a portion of the larger W43 massive star-forming region studied by \citet{Saral_2015}. Therefore, we decided to study the possible variability of the YSOs associated with IRAS 18456-0223, and examine the properties of its stellar population chiefly using archival infrared
databases. If the flaring star was detected in any of these surveys, which were conducted decades after POSS, then it would be most likely an FUor. It would also be possible to fit a spectral energy distribution for a better understanding of its nature.

Our study is based on the data discussed in Sec ~\ref{sec:data}. In Sec~\ref{sec:YSO population}, we discuss identification and characteristion of the YSO population. Distance estimates to the region are discussed in Sec~\ref{sec:distance_estimate}. Spatial distribution of the YSOs and overall properties of the cloud are discussed in Sec~\ref{sec:spatial distribution}. In Sec~\ref{sec:discussions}, we describe deductions from our optical spectra and put our work in context of previous studies. Salient points of this work are covered in the concluding part, Sec~\ref{sec:conclusions}.

\section{OBSERVATIONS AND DATA REDUCTION}
\label{sec:data}

\subsection{Archival datasets and completeness}

\label{sec:data2}
We utilized archival multiwavelength data from optical, near-infrared (NIR), mid-infrared (MIR), and far-infrared (FIR) bands within a \(10~\arcmin \times 10~\arcmin\) region centered on IRAS~18456$-$0223. This field of view was adopted because larger areas begin to overlap with neighbouring IRAS sources, and a visual comparison with the cluster extent presented in \citet{Saral_2015} confirms that the embedded cluster is fully contained within this region. This choice also provides coverage well beyond the \(4~\arcmin \times 4~\arcmin\)field used by \citet{2001A&A...370..991K}, ensuring a conservative and complete sampling of the cluster environment. The details of catalogues used are summarized in the following paragraphs.

We used astrometric and photometric data from \textit{Gaia DR3} (see \citet{2023A&A...674A..41G}), including source positions, parallaxes, and proper motions, along with magnitudes in the $G$ (330--1050\,nm), $G_{\mathrm{BP}}$ (330--680\,nm) and, $G_{\mathrm{RP}}$ (640--1050\,nm) bands \citep[see for details,][]{2016A&A...595A...2G,2023A&A...674A..32B}.

NIR $JHK_{\mathrm{s}}$ data were obtained from the Two Micron All-Sky Survey (\textit{2MASS}) Point Source Catalog (PSC) \citep{2006AJ....131.1163S}. Sources having uncertainties less than 0.1 mag in all three $J$ (1.25\,$\mu$m), $H$ (1.65\,$\mu$m), and $K_{\mathrm{s}}$ (2.17\,$\mu$m) bands were selected to ensure high-quality data. We also used the photometry in the $J$, $H$ and, $K$, obtained from the Galactic Clusters Survey (GCS) of the UKIRT Infrared Deep Sky Survey (\textit{UKIDSS}) Release 11 \citep{2008MNRAS.391..136L}.

MIR data of this region were obtained using the \textit{Spitzer} Space Telescope as part of the Galactic Legacy Infrared Midplane Survey Extraordinaire (GLIMPSE) programme \citep{2003PASP..115..953B,2009PASP..121..213C}. The data were retrieved from the InfraRed Science Archive (Spring ’07 Archive: more complete catalogue) and the APOGLIMPSE Catalog and Archive (more complete catalogue). GLIMPSE observations were carried out using the InfraRed Array Camera (IRAC) in the 3.6, 4.5, 5.8, and 8.0\,$\mu$m bands. We also make use of AllWISE processing of the
WISE survey data \citet{2019ipac.data...I1W}. 

We obtained far‐infrared (FIR) continuum emission data from the \textit{\textit{Herschel}} Space Observatory \citep{2010A&A...518L...1P} using its two primary instruments: the Photodetector Array Camera and Spectrometer (PACS) \citep{2010A&A...518L...2P} and the Spectral and Photometric Imaging Receiver (SPIRE) \citep{2010A&A...518L...3G}. PACS has $6\arcsec$ resolution at 70 $\mu$m and 100 $\mu$m, and approximately $12\arcsec$ at 160 $\mu$m. SPIRE provides spatial resolutions of $18.1\arcsec$, $25.2\arcsec$, and $36.9\arcsec$ at 250 $\mu$m, 350 $\mu$m, and 500 $\mu$m, respectively. These publicly available data were retrieved from the \textit{Herschel} Science Archive and accessed via the IRSA/IPAC Image Service under the \textit{Herschel} High Level Images (HHLI).

The \textit{Gaia DR3} catalogue provides high-precision astrometry and photometry with a limiting magnitude of $G \sim 20.7$ mag. The \textit{2MASS} PSC is complete down to $J \sim 15.8$, $H \sim 15.1$, and $K_{\mathrm{s}} \sim 14.3$ mag. To ensure high data quality in \textit{UKIDSS} and keep only stellar sources, the photometric errors for all three bands were taken to be less than 0.2 mag and values of the ‘PSTAR’ flag larger than 0.9. This resulted in limiting magnitudes of 20.9, 19.4, and 19.0 mag for the $J$, $H$, and $K$ bands, respectively. The MIR \textit{Spitzer}/GLIMPSE data reach typical limiting magnitudes of $\sim$15.5, 15.0, 13.0, and 12.0 mag in the 3.6, 4.5, 5.8, and 8.0\,$\mu$m bands, while the \textit{WISE} AllWISE catalogue is sensitive to $\sim$16.5, 15.5, 11.2, and 7.9 mag in the W1--W4 bands; however, completeness in the MIR is affected by spatially varying background emission and source confusion. At FIR wavelengths, the \textit{Herschel} PACS and SPIRE data are limited by angular resolution and are therefore incomplete for resolving individual YSOs.\textbf{
}
\subsection{Optical Spectroscopy }
We carried out spectroscopic observations of the three bright sources near the flaring star using the Hanle
Faint Object Spectrograph Camera (HFOSC) \footnote{\url{https://www.iiap.res.in/centers/iao/facilities/HCT/hfosc/}} mounted on the 2m
Himalayan Chandra Telescope (\textit{HCT}) of the Indian Astronomical
Observatory (IAO) on 25 August 2007.  The aim was to test the possibility that these stars were also YSOs and related to the flaring star because of their angular proximity (see Fig. 3 of \cite{2001A&A...370..991K}).
Feige 110 was observed as the standard star. All the exposures were for 600 s. 
We used a combination of the 167l slit (1.92\arcsec
× 11\arcmin) and grism 8, covering the wavelength range of 5300 to 9100 \AA, resulting in a spectral resolution of $\sim 3$\,\AA.  We used FeNe lamp spectra for wavelength calibration and night sky lines for further small correctional shifts.  All preprocessing and data reductions were performed in a standard manner using various tasks available in IRAF.

\section{IDENTIFICATION OF YOUNG STELLAR OBJECTS}
\label{sec:YSO population}
\begin{figure*}[t!]
    \centering

   \begin{subfigure}{0.33\textwidth}
        \centering
        \includegraphics[width=\textwidth, trim=0.5cm 0cm 0cm 0.5cm, clip]{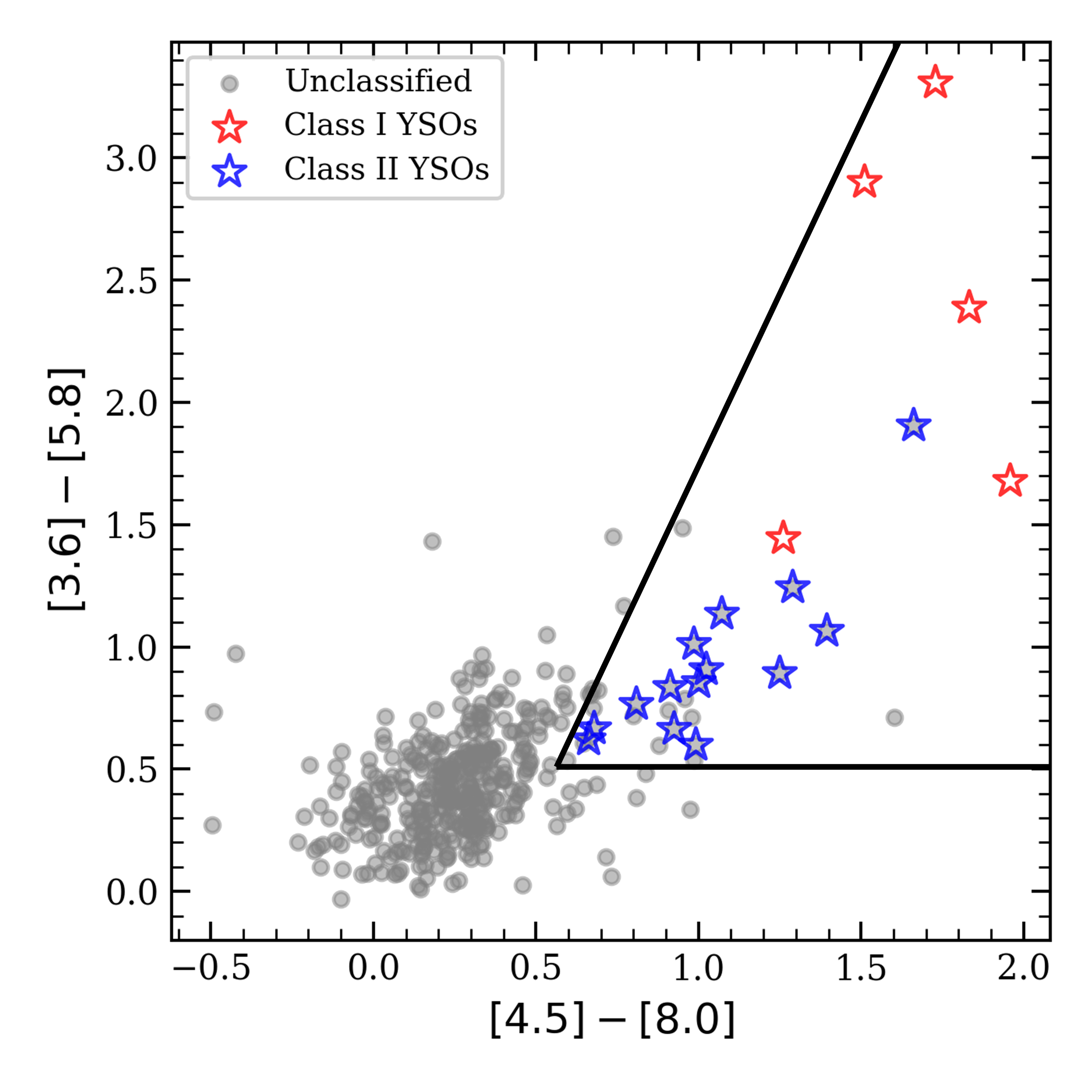}
    \end{subfigure}
    \hspace{0.5cm}
    \begin{subfigure}{0.33\textwidth}
        \centering
        \includegraphics[width=\textwidth, trim=0.5cm 0cm 0cm 0.5cm, clip]{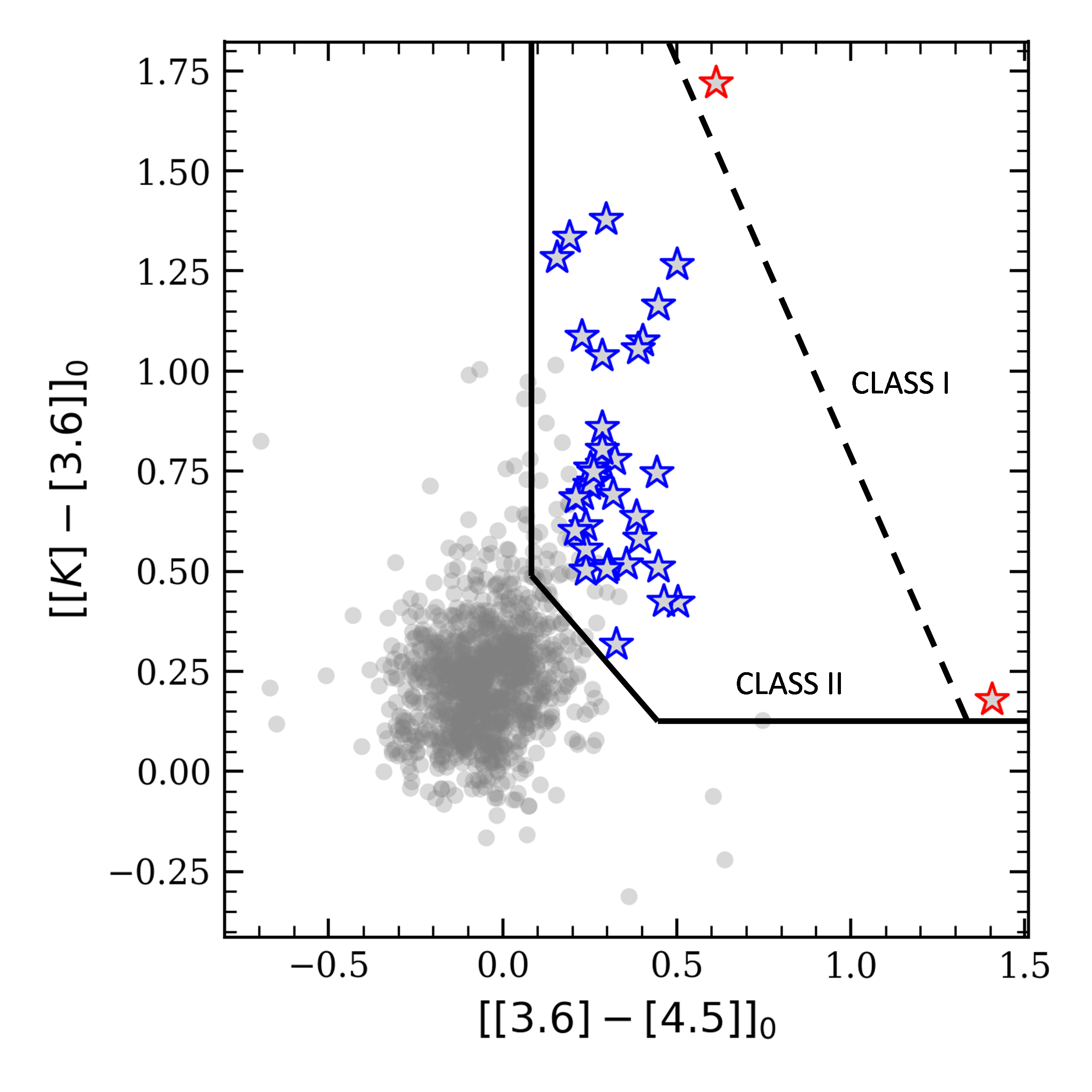}
    \end{subfigure}

    \vspace{0.1cm} 
    \begin{subfigure}{0.33\textwidth}
        \centering
        \includegraphics[width=\textwidth, trim=0.5cm 0cm 0cm 0.5cm, clip]{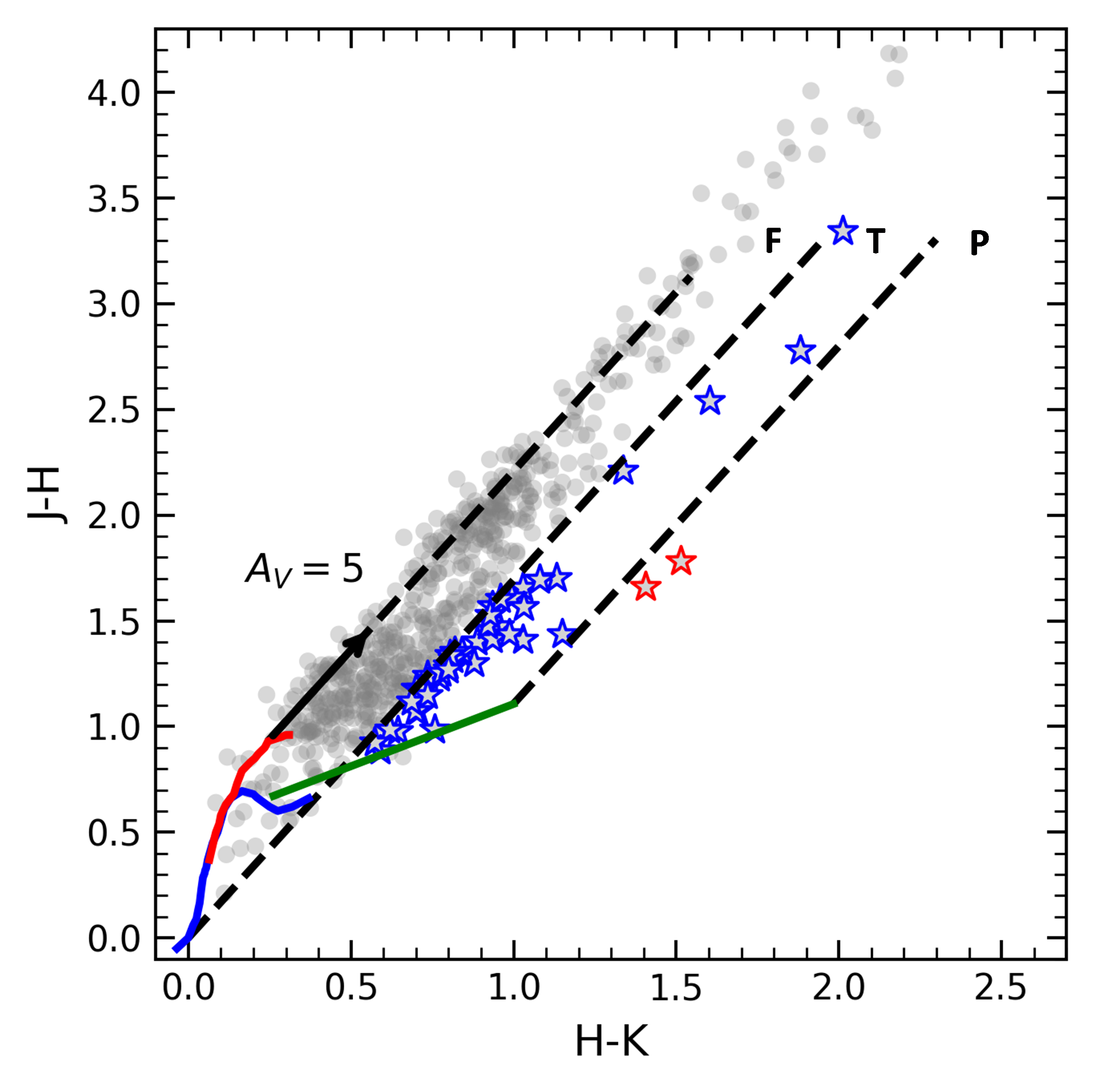}
    \end{subfigure}
    \hspace{0.5cm}
    \begin{subfigure}{0.33\textwidth}
        \centering
        \raisebox{0.2cm}{\includegraphics[width=\textwidth, trim=0cm 0.25cm 0cm 0cm, clip]{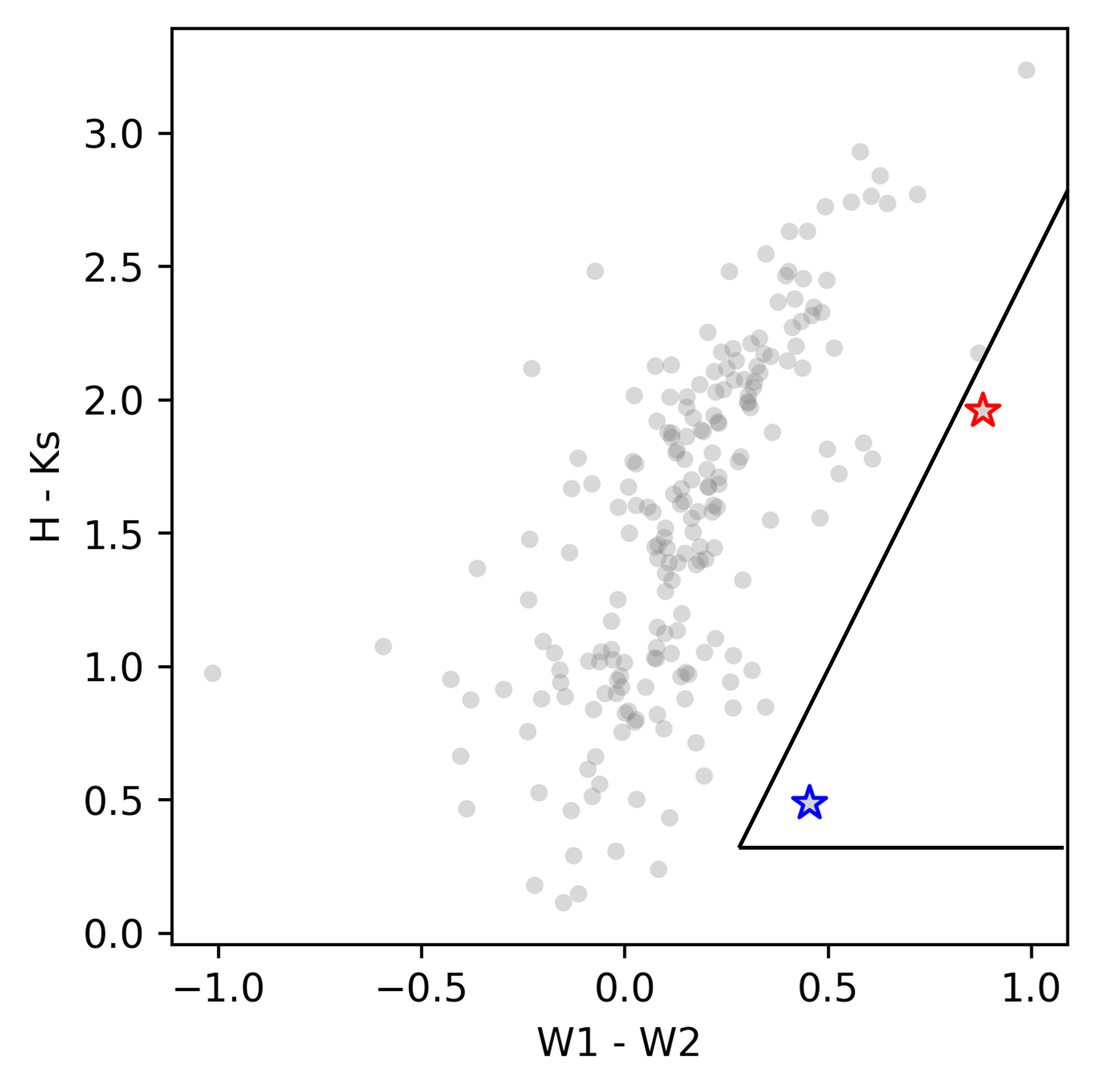}}
    \end{subfigure}
    \caption{ (a) MIR CCD using the procedure outlined in \citet{2009ApJS..184...18G}. class I and class II are highlighted in red and blue, respectively.  
    (b) NIR–MIR CCD using the procedure of \citet{2009ApJS..184...18G}. The regions of class I and class II sources have been labelled.  
    (c) NIR CCD in the CIT system. The red, blue curve, and green line show the giant locus, dwarf locus \citep{1988PASP..100.1134B}, and the CTTS locus \citep{1997AJ....114..288M}, respectively. Three parallel slanted dashed lines mark the reddening vectors, drawn using the extinction law of \citet{1981ApJ...249..481C}. The reddening lines extending from the tip of the giant branch to the base of the main-sequence dwarf locus delineate the main-sequence reddening band. Sources located within the region labeled `F' are likely field stars or evolved cluster members with little or no near-infrared excess. The sources in `T' and `P' regions are taken to be class II and class I sources respectively. However, we there may be overlap of the Herbig Ae/Be stars
in the “T” region \citep{hillenbrand2002youngcircumstellardisksevolution}.(d) NIR-MIR CCD by \citet{2014ApJ...791..131K}}
    \label{Fig2a}%
\end{figure*}

The following steps were followed for the identification and classification of Young Stellar Objects (YSOs):
\begin{enumerate}
\item{}
We applied the Phase I identification scheme described in \citet{2009ApJS..184...18G} in Spring ’07
archive for data present in all four IRAC bands 3.6, 4.5, 5.8, and 8.0 $\mu$m having photometric uncertainties $\sigma$ $\leq$0.2 mag. `[4.5] - [8.0]' versus `[3.6] - [5.8]' colour colour diagram (CCD) as shown in Fig.\,\ref{Fig2a}(a). We identified 5 class I and 14 class II with this scheme.

\item{}
All sources need not have detections at 5.8~$\mu$m and 8.0~$\mu$m; they may instead be well detected in the near-infrared (NIR) bands.  To select YSO candidates among such objects, we follow the procedure of \citet{2009ApJS..184...18G}, applied to the APOGlimpse catalogue, using the $H$, $K_{\mathrm{s}}$, 3.6~$\mu$m and 4.5~$\mu$m bands. For high quality data we further set that the photometric uncertainties satisfy 
3.6 and 4.5 $\mu$m  $\leq$0.2
 and $\leq$0.1 for $H$ and $K_{\mathrm{s}}$. The YSOs were identified from their location in the dereddened, using the colour excess ratios from \citet{2007ApJ...663.1069F}, the dereddening process is defined in \citet{2009ApJS..184...18G}. The corresponding plot ‘$K_{\mathrm{s}}$–$[3.6]$’ versus ‘$[3.6]$–$[4.5]$’ CCD, is shown in Fig.\,\ref{Fig2a}(b) with 2 class I and 34 class II sources identified.

\item{}
Young disk-bearing stars exhibit near-infrared (NIR) excess emission, which can be detected through their position in the NIR CCD.  We utilized the ($J$-$H$) vs. ($H$-$K_{\mathrm{s}}$) CC diagrams as shown in Fig.\,\ref{Fig2a}(c) to identify stars with infrared (IR) excess in our regions of interest. We applied the extinction laws from \citet{1981ApJ...249..481C}, with values of $A_J/A_V = 0.265$, $A_H/A_V = 0.155$, and $A_K/A_V = 0.090$
to the sources themselves. Based on the NIR CC diagram, sources are categorized into three regions: F, T, and P, as described by \citet{Ojha_2004a, 2007MNRAS.380.1141S, 2008MNRAS.383.1241P}. We identified 2 class I and 38 class II sources.

\begin{table*}[htbp]
  \centering
  \caption{Photometric magnitudes (with uncertainties) and classification for selected sources.\label{tab:photometry}}
  \renewcommand{\arraystretch}{1.1}
  \resizebox{\textwidth}{!}{%
  \begin{tabular}{crrlccccccc}
    \hline\hline
    RA (deg) & Dec (deg) & Class & Method & $J$ & $H$ & $K_{\mathrm{s}}$ & [3.6] & [4.5] & [5.8] & [8.0] \\
    \hline
    \multicolumn{11}{c}{\it Selected YSOs} \\
    \hline
    281.9845 & -2.2869 & II & a   & 15.535 $\pm$ 0.076 & 14.619 $\pm$ 0.074 & 13.985 $\pm$ 0.093 & 13.721 $\pm$ 0.097 & 13.530 $\pm$ 0.166 & \ldots & \ldots \\
    281.9864 & -2.2873 & II & b  & 16.993 $\pm$ \ldots & 15.163 $\pm$ 0.074 & 13.814 $\pm$ 0.055 & 12.282 $\pm$ 0.045 & 11.755 $\pm$ 0.034 & \ldots & \ldots \\
    281.9874 & -2.2569 & II & c & 12.741 $\pm$ 0.026 & 11.598 $\pm$ 0.025 & 11.108 $\pm$ 0.027 & \ldots & \ldots & \ldots & \ldots \\
    281.9937 & -2.3273 & II & b  & 16.341 $\pm$ \ldots & 15.345 $\pm$ 0.099 & 14.295 $\pm$ 0.087 & 13.447 $\pm$ 0.080 & 12.982 $\pm$ 0.075 & \ldots & \ldots \\
    281.9966 & -2.2760 & II & b  & 16.691 $\pm$ \ldots & 15.242 $\pm$ \ldots & 14.378 $\pm$ 0.089 & 12.090 $\pm$ 0.068 & 11.632 $\pm$ 0.085 & 10.955 $\pm$ 0.085 & 10.560 $\pm$ 0.101 \\
    \hline
  \end{tabular}
  }
  \vspace{2mm}
  \begin{minipage}{\textwidth}
    \footnotesize
    \textbf{Notes.} \\
    \textit{Method column:} a – \textit{2MASS} colour classification; b – Gutermuth et al. (2009) \textit{Spitzer}-based classification; c – Koenig et al. (2014) \textit{WISE}-based classification. \\
    Table~\ref{tab:photometry} is available in its entirety in a machine-readable form in the online journal. A portion is shown here for guidance regarding its form and content.
  \end{minipage}
\end{table*}

\item{}
Additional YSOs were identified from the \textit{ALLWISE} catalog using \textit{WISE} MIR and 2MASS NIR data. 
We followed the alternate procedure described by \citet{2014ApJ...791..131K}, which modifies the K12 scheme and uses the 
$H - K_{\mathrm{s}}$ versus $W1 - W2$ colour--colour diagram to identify and classify YSOs,
because of the limited sensitivity of the last two \textit{WISE} band. We identified 1 class I and 1 class II source, and the corresponding colour-colour diagram is shown in Fig.\,\ref{Fig2a}(d)

\end{enumerate}
Finally, we combined the sources identified across all steps with matching radius of 4 decimal places to construct a single YSO master list (see Table~\ref{tab:photometry}). In cases where the same source appeared in multiple steps with different classifications, we assigned the class corresponding to the earliest step in which it was identified. In total, we obtained 9 Class~I and 80 Class~II sources within a field of view of \(10~\arcmin \times 10~\arcmin\) encompassing the molecular cloud. Table~\ref{tab:photometry} presents a subset of the final YSO catalogue.
The flaring star was not detected in any of the above surveys. Hence, nothing can be said about its behaviour in the infrared.

\section{DISTANCE TO IRAS 18456-0223}
\label{sec:distance_estimate}
\subsection{Membership analysis and distance using \textit{Gaia DR3}}

Stars in a cluster share similar kinematic properties and mean motions. Therefore, proper motions of stars are crucial for identifying cluster members among field stars \citep{10.1093/mnras/stt136,Bisht_2022}. We used data from {\it Gaia} DR3 to estimate the membership probability of stars and the distance to the cluster IRAS 18456-0223.

\begin{figure*}[h!]
    \centering
    \hspace{-0.5cm}
    \includegraphics[width=0.95\hsize, trim=1.5cm 2.5cm 0cm 2.2cm, clip]{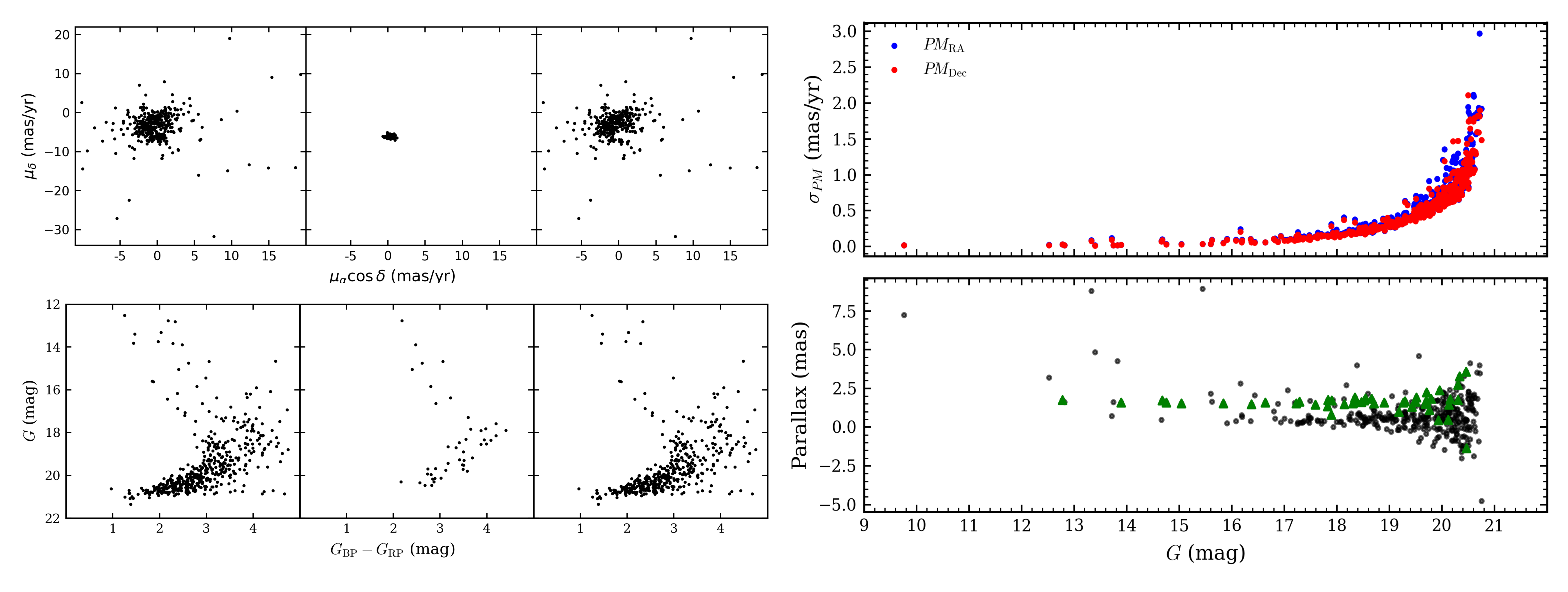} %
\caption{
\textbf{Left panel:} Proper motion vector-point diagrams (VPDs; top subpanels) and \textit{Gaia DR3} $G$ vs. $(G_\mathrm{BP} - G_\mathrm{RP})$ colour–magnitude diagrams (CMDs; bottom subpanels) for stars in the IRAS 18456-0223 region. The leftmost subpanels show all stars, while the middle and right subpanels display the high-probability cluster members and field stars, respectively. 
\textbf{Right panel:} Proper motion uncertainties ($\sigma_\mathrm{PM}$) and parallaxes plotted as a function of $G$ magnitude. High-probability members with $P_\mu > 50\%$ are indicated by green triangles.}

    \label{fig:gaia_plot}
\end{figure*}

A vector point diagram (VPD) was constructed by plotting the proper motion components $\mu_{\alpha} \cos \delta$ (mas yr$^{-1}$) and $\mu_{\delta}$ (mas yr$^{-1}$), as shown in the left panel of Fig.~\ref{fig:gaia_plot}. Each point in the VPD represents the proper motion of a star. The dense region in the VPD indicates stars that share similar motion, a key criterion for cluster membership. An eye-estimated region was selected around this dense area to select preliminary members of the cluster. 

To calculate the membership probability of these stars, we followed the method of \citet{1998A&AS..133..387B}, as described in \citet{Bisht_2020}. This method uses two distribution functions: $\phi_{\nu c}$ for cluster stars and $\phi_{\nu f}$ for field stars.

From the fit to the VPD by the preliminary members, the cluster’s proper-motion centroid was chosen at $\mu_{x,c} = 0.45$ mas yr$^{-1}$ and $\mu_{y,c} = -6.01$ mas yr$^{-1}$, with dispersions $\sigma_{x,c} = 0.50$ mas yr$^{-1}$ and $\sigma_{y,c} = 0.51$ mas yr$^{-1}$. In contrast, the field population has $\mu_{x,f} = -0.350$ mas yr$^{-1}$, $\mu_{y,f} = -3.53$ mas yr$^{-1}$, $\sigma_{x,f} = 4.42$ mas yr$^{-1}$, and $\sigma_{y,f} = 6.13$ mas yr$^{-1}$. Adopting a dispersion for the cluster of $\sigma_c = 0.35$ mas yr$^{-1}$ \citep{1989AJ.....98..227G}, the membership probability for each star $i$ was calculated,\begin{equation}
P_{\mu}(i) = \frac{n_c \times \phi_{\nu_c}(i)}{n_c \times \phi_{\nu_c}(i) + n_f \times \phi_{\nu_f}(i)}
\label{eq:membership_probability}
\end{equation}
where $n_c$ and $n_f$ are the normalized number of stars in the cluster and field regions, respectively. Based on this, we identified 47 stars with $P_\mu > 50\%$ as high probable cluster members (Fig~\ref{fig:gaia_plot}).

To determine the mean proper motion of IRAS 18456-0223, histograms were plotted for $\mu_{\alpha} \cos \delta$ and $\mu_{\delta}$ using these members. Fitting Gaussian functions yielded $\mu_{\alpha} \cos \delta = 0.4533 \pm 0.50$ mas yr$^{-1}$ and $\mu_{\delta} = -6.017 \pm 0.51$ mas yr$^{-1}$.

\begin{figure}[h!]
   \centering
   
   \includegraphics[width=0.85\hsize]{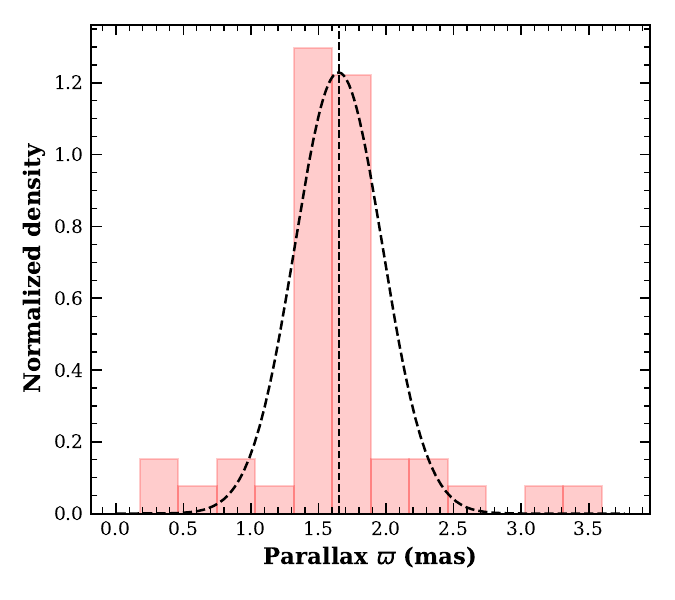}
      \caption{Histogram of parallax values for cluster members, with a Gaussian profile overlaid. The best-fit Gaussian has a mean parallax of \(1.649 \pm 0.325\)~mas. Only sources with positive parallaxes were considered, and the fit was restricted to the range \(0.5 < \varpi < 3\)~mas to exclude outliers.}
         \label{figplx}
   \end{figure}

We also plotted a histogram of the parallax values of cluster members, rejecting spurious stars with negative parallaxes. This distribution was fitted with a Gaussian profile (Fig.~\ref{figplx}), yielding a mean parallax of $1.649 \pm 0.325$ mas. This corresponds to a heliocentric distance of $606 \pm 119.5$ pc.

\subsection{Distance based on the known YSO candidates associated
with the cloud}
\begin{figure}[h!]
   \centering
   
   \includegraphics[width=0.85\hsize]{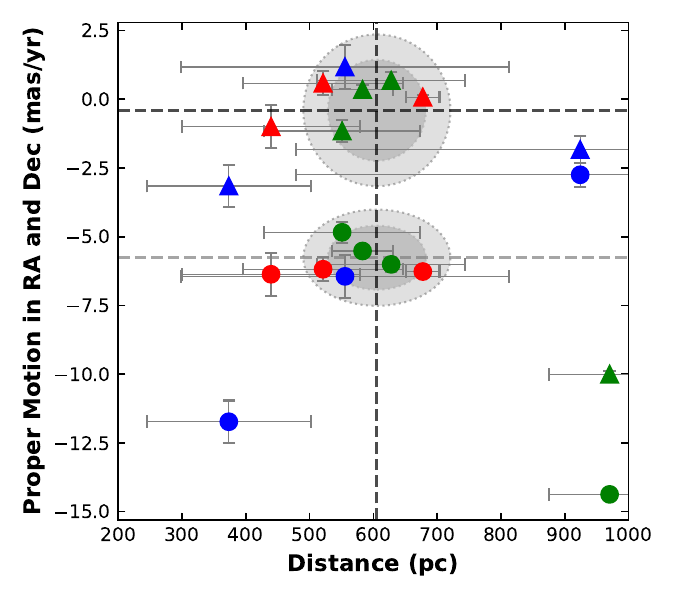}
      \caption{Proper motion versus distance for \textit{Gaia DR3} YSO candidates. Triangles denote $\mu_{\alpha}$ and circles denote $\mu_{\delta}$, with error bars representing uncertainties in both distance and proper motion. Dashed lines indicate the combined medians for the green sources, and overlaid error ellipses (2$\times$MAD in a darker shade and 3$\times$MAD in a lighter shade) illustrate the dispersion.}
         \label{fig2}
   \end{figure}

One of the direct ways of estimating distances to a molecular cloud is to use the stars associated with the cloud as described in \citet{2020MNRAS.494.5851S}. Among our 89 YSO candidates, we could determine distance and proper motion values in right ascension ($\ mu_\alpha$) and declination ($\mu_\delta$) for 42 sources using \textit{Gaia DR3}, with 10 sources having the ratio $m/\sigma_m$ (where $m$ represents the distance, $\mu_\alpha$, and $\mu_\delta$ values, and $\sigma_m$ their respective errors) $\geq 1$. The \textit{Gaia DR3} results for the YSO candidates are shown in Fig.\ref{fig2}, where triangles and circles represent distance-$\mu_\alpha$ and distance-$\mu_\delta$ values.  SNR is defined as the ratio of the measurement to its uncertainty (i.e., SNR$_d = d/e_d$, SNR$_{\mu_{\alpha}} = |\mu_{\alpha}|/e_{\mu_{\alpha}}$, SNR$_{\mu_{\delta}} = |\mu_{\delta}|/e_{\mu_{\delta}}$) and is used to classify sources into three groups: green (SNR$_d\ge3$, SNR$_{\mu_{\alpha}}\ge2$, SNR$_{\mu_{\delta}}\ge2$), red (SNR$_d\ge3$, $1\le$SNR$_{\mu_{\alpha}}<$2, SNR$_{\mu_{\delta}}\ge2$), and blue (2$\le$SNR$_d<$3, SNR$_{\mu_{\alpha}}\ge1$, SNR$_{\mu_{\delta}}\ge2$). Among these 10 sources, 3 (red circles and triangles), 3 (blue circles and triangles), and 4 (green circles and triangles) were classified based on their SNR. In this study, we selected only high-quality (green) sources with distances less than one kpc.

\begin{figure}[htbp!]
   \centering
   
   \includegraphics[width=0.8\hsize]{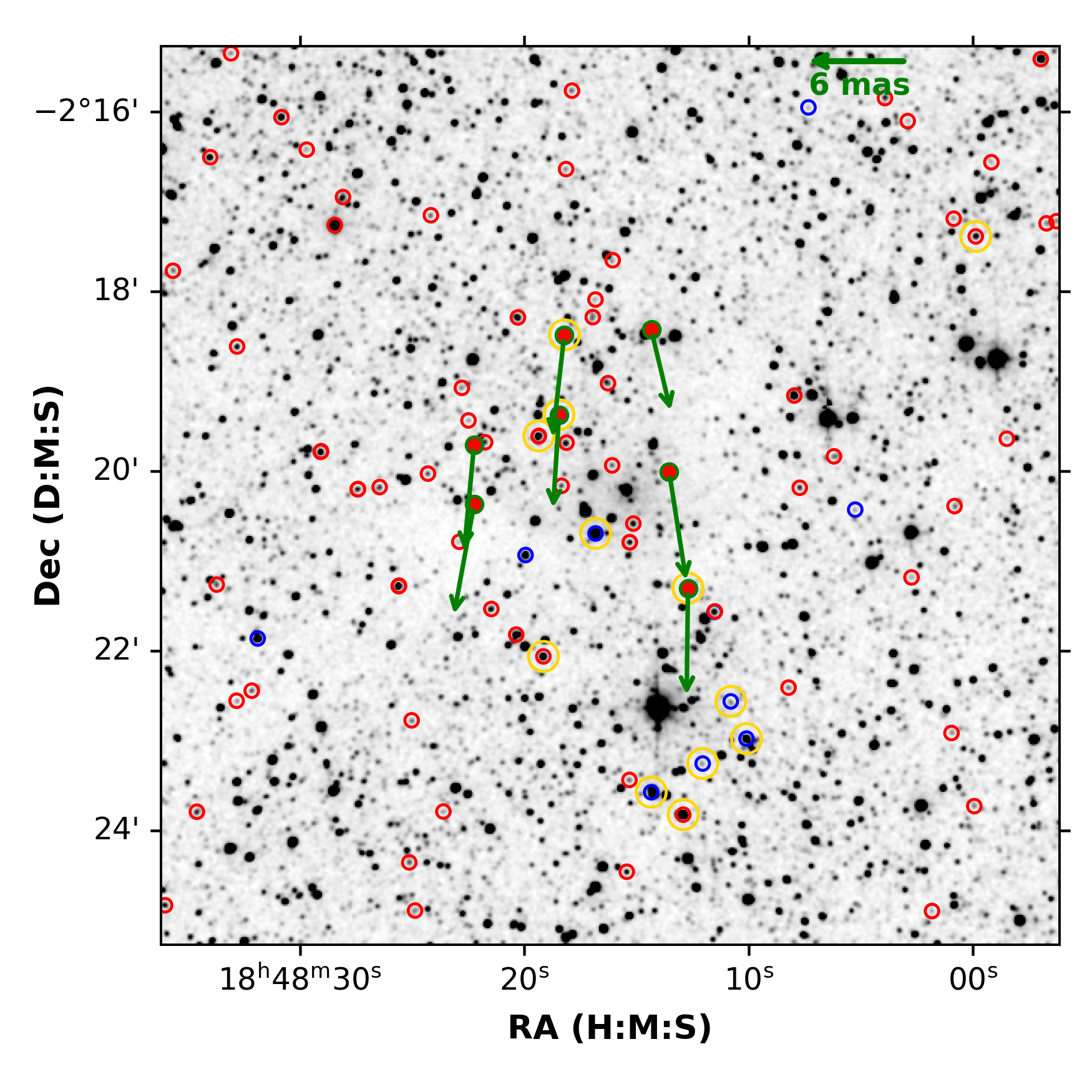}
      \caption{Proper Motion Plot for the YSOs (green and red sources taken, green arrows) overlaid in Ks band Image obtained from the \textit{2MASS} Interactive Image Service at IRSA. The distribution of YSOs based on their class is also depicted where red sources are class II while blue sources are class I. Yellow-marked sources were successfully fitted using SED models.}
         \label{fig3}
   \end{figure}

Based on the proper motion and distance values, a clustering of sources is evident in Fig.\ref{fig2}. The medians of $\mu_\alpha$, $\mu_\delta$, and distance are -0.40, $-5.76\,\text{mas yr}^{-1}$, and 605 pc, respectively. The variance and standard deviation are commonly used to measure data spread, but are sensitive to extreme values. Therefore, we use the median absolute deviation (MAD) to estimate statistical dispersion, as it is more robust to outliers. The MAD values for distance, $\mu_\alpha$, and $\mu_\delta$ are 38.44 pc, 0.92, and 0.58 mas yr$^{-1}$, respectively. Six YSO candidates fall within three times the MAD in proper motions and distances, as shown in Fig.\ref{fig2} within the lighter gray-shaded ellipses. The remaining four sources exhibit deviations from the median value. The proper motion of YSOs is illustrated in Fig.~\ref{fig3}.

Finally, an average sampling of data from both the results (all stars and only YSOs) yields an estimated distance of \textbf{$606 \pm 110\mathrm{pc}$}.

 The flaring star does not have a Gaia counterpart, hence its distance and, consequently, its luminosity is unknown. The POSS detection by \cite{2001A&A...370..991K} is the only data point available for this star. Due to the absence of a lightcurve, its true nature as a possible FUor or EXor remains undetermined.

\section{CHARACTERISTICS OF YSOs}

\subsection{SED of YSOs}
\label{subsection:SEDS_of_YSOs}
To estimate the physical parameters of selected young stellar objects (YSOs), we performed Spectral Energy Distribution (SED) modeling using the grid of YSO models developed by \citet{2006ApJS..167..256R} and and utilized the corresponding online SED–fitting interface described in \citet{2007ApJS..169..328R}. This extensive modeling employed the radiative transfer code described in \citet{2003ApJ...598.1079W,2003ApJ...591.1049W}. The online fitting tool matches these models to observed data, evaluating each fit using a chi-squared ($\chi^2$) parameter to quantify the goodness of fit. The distance and the interstellar visual extinction ($A_V$) are treated as free parameters, with the distance range specified from 0.4 to 0.8~kpc based on the uncertainties involved (see Sec.\ref{sec:distance_estimate} for the distance estimates). The extinction range was constrained using sources in the same field that do not 
satisfy the YSO selection criteria (non-YSO sources) and are therefore predominantly foreground and 
background field stars (see \citet{2015MNRAS.447.2307M}). Almost all of these sources show  $A_V \lesssim 20$~mag, providing an empirical upper limit to the line-of-sight 
interstellar extinction toward the region. While YSOs may experience additional 
local extinction due to their natal environments, unphysically large foreground 
extinction values are not expected. We therefore adopted a conservative 
extinction range of $1 \leq A_V \leq 20$~mag for the SED fitting.

Given the extensive parameter space of the model grid, it was essential to have comprehensive photometric coverage across various wavelengths to effectively constrain the SED fits. Therefore, we selected only those YSOs with available photometry in the $J$, $H$, and $K_{\mathrm{s}}$ bands, all four IRAC bands, and \textit{WISE} bands 3 and 4, resulting in a sample of 12 sources. To account for potential underestimations in flux uncertainties and to ensure unbiased fitting, we assigned a uniform photometric uncertainty of 10\% to both near-NIR and MIR data. The SEDFITTER tool evaluates the quality of each fit using the chi-squared ($\chi^2$) statistic. Following the methodology outlined by \citet{2007ApJS..169..328R}, we considered models as acceptable fits if they satisfied the condition $\chi^2 - \chi^2_{\rm min} < 3$ per data point, where $\chi^2_{\rm min}$ represents the minimum chi-squared value among all models. This criterion allowed us to identify a subset of well-fitting models for each YSO, providing insights into their physical characteristics.

\begin{figure}[htbp!]
    \centering
    
    \includegraphics[width=0.88\hsize]{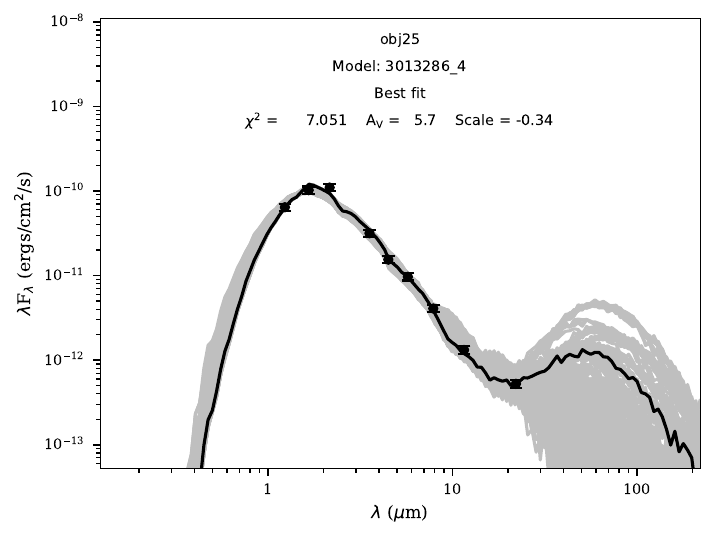}
    \caption{SED fitting using the online tool of \citet{2007ApJS..169..328R} for a class 2 source in Table~\ref{tab:sed_errors}. The black dots mark the observed data points. The solid black curve represents the best-fitting model, while the grey curves indicate subsequent good fits for which $\chi^2 - \chi^2_{\text{min}}$ (per data point) $< 3$.
}
    \label{fig:exsed}
\end{figure}

Table\,\ref{tab:sed_errors}. summarizes the SED modeling results, listing the physical parameters such as the age of the central source ($T$), stellar mass, disc mass ($M_{\rm disc}$), disc accretion rate ($\dot{M}_{\rm disc}$), effective temperature of the central source, total system luminosity ($L_{\rm total}$), interstellar visual extinction ($A_V$), and the minimum $\chi^2$ per data point. The SED results cover 5 class I, 7 class II sources (See example, Fig.\ref{fig:exsed}). Although the number of YSOs is modest, these results still provide insight into the physical properties of the forming stellar sources in this region. As shown in Table\,\ref{tab:sed_errors}, there is a notable age dispersion, with most sources ranging from approximately 0.002 to 4~Myr, while 3 YSOs exhibit ages greater than 1~Myr, which suggests ongoing star formation. All analyzed YSOs have masses greater than 0.1~$M_\odot$, and four exceed 2~$M_\odot$. It is important to note that the SED results represent approximations of the true values, as a model that provides an empirically consistent fit might not necessarily be the correct one. For example, the SED fitting results in a statistically lower age for class II YSOs than those of class I, while the opposite is typically expected.

The SED fitting framework of \citet{2017A&A...600A..11R}  employs a revised model grid in which only parameters that directly affect radiative transfer are fitted, while quantities such as stellar mass and age are excluded due to their dependence on evolutionary tracks. In contrast, the \citet{2006ApJS..167..256R} models provide estimates of stellar mass and age. Since a primary objective of this work is to estimate stellar masses and ages and to enable direct comparison with earlier results, the R06 framework is adopted, and the model-dependent nature of the derived stellar parameters is explicitly acknowledged.


\subsection{Variability of YSOs}
88 of the 89 YSO and T Tauri candidates identified by us have a \textit{2MASS} and \textit{UKIDSS} counterpart within 1\arcsec. Since YSOs embedded in substantial amounts of circumstellar gas and dust are expected to exhibit significant photometric variability \citep{1997A&A...320L..41H}, we therefore searched for signatures of photometric variability among the YSOs in this region. For this purpose, we calculated the $\Delta J$ and $\Delta K_{\mathrm{s}}$ magnitudes as the difference between the \textit{UKIDSS} magnitudes converted into the \textit{2MASS} system \citep{2006MNRAS.367..454H} and the \textit{2MASS} magnitudes.  

Nine sources show variations greater than 0.5~mag in the $J$ band, and eight sources vary by more than 0.5~mag in the $K_{\mathrm{s}}$ band over a timescale of six years (\textit{2MASS}: 11 April 1999; \textit{UKIDSS}: 22 June 2005), strengthening their identification as young stars.
Since the flaring star does not have counterparts in 2MASS or UKIDSS, there is no way of knowing whether it was variable during this time interval.  


\begin{table*}
  \caption{SED best-fit parameters with 1$\sigma$ errors (Age in Myr)}
  \label{tab:sed_errors}
  \centering
  \resizebox{\textwidth}{!}{%
    \begin{tabular}{cccccccccc}
      \toprule
      RA (deg) & Dec (deg)
        & $A_V$ (mag)
        & Age (Myr)
        & Mass ($M_\odot$)
        & $T_{\mathrm{eff}}$ (K)
        & $M_{\mathrm{disk}}$ ($M_\odot$)
        & $\dot M_{\mathrm{disk}}$ ($M_\odot\,$yr$^{-1}$)
        & $L_{\mathrm{tot}}$ ($L_\odot$) \\
      \midrule
      \multicolumn{9}{c}{\bfseries class I} \\
      \cmidrule(lr){1-9}
      282.07011 & $-2.34487$ & $3.94\pm1.21$ & $2.78\pm0.70$ & $7.22\pm1.03$ & $20980\pm2070$ & $1.63\times10^{-2}\pm1.26\times10^{-2}$ & $1.20\times10^{-5}\pm5.95\times10^{-6}$ & $2.54\times10^{3}\pm9.41\times10^{2}$ \\
      282.05978 & $-2.39289$ & $6.07\pm0.00$ & $0.14\pm0.00$ & $1.10\pm0.00$ & $4167\pm0.00$       & $7.60\times10^{-3}\pm0.00$             & $4.61\times10^{-8}\pm0.00$             & $1.24\times10^{1}\pm0.00$             \\
      282.05021 & $-2.38757$ & $14.63\pm0.00$& $4.06\pm0.00$ & $2.53\pm0.00$ & $9462\pm0.00$       & $3.84\times10^{-3}\pm0.00$             & $4.99\times10^{-8}\pm0.00$             & $5.73\times10^{1}\pm0.00$             \\
      282.04500 & $-2.37604$ & $1.00\pm3.15$ & $0.07\pm1.95$ & $3.74\pm0.06$ & $4430\pm4790$    & $7.00\times10^{-2}\pm2.96\times10^{-2}$& $7.42\times10^{-7}\pm3.68\times10^{-7}$& $1.07\times10^{2}\pm3.79\times10^{1}$\\
      282.04208 & $-2.38294$ & $2.58\pm0.00$ & $0.29\pm0.00$ & $1.41\pm0.00$ & $4379\pm0.00$       & $6.95\times10^{-4}\pm0.00$             & $8.53\times10^{-10}\pm0.00$            & $9.76\times10^{0}\pm0.00$             \\
      \midrule
      \multicolumn{9}{c}{\bfseries class II} \\
      \cmidrule(lr){1-9}
      282.08063 & $-2.32676$ & $4.07\pm0.67$ & $0.35\pm0.04$ & $2.65\pm0.63$ & $4678\pm163$     & $6.12\times10^{-2}\pm2.48\times10^{-2}$& $2.46\times10^{-7}\pm1.22\times10^{-7}$& $2.13\times10^{1}\pm5.12\times10^{0}$\\
      282.07979 & $-2.36768$ & $10.75\pm0.72$& $2.48\pm0.93$ & $1.24\pm1.02$ & $4423\pm846$     & $1.21\times10^{-3}\pm2.63\times10^{-3}$& $2.39\times10^{-10}\pm1.61\times10^{-9}$& $1.43\times10^{0}\pm2.84\times10^{0}$\\
      282.07689 & $-2.32283$ & $4.91\pm1.19$ & $0.00\pm0.02$ & $0.22\pm0.06$ & $3118\pm172$     & $2.33\times10^{-4}\pm1.60\times10^{-3}$& $4.55\times10^{-9}\pm8.24\times10^{-9}$ & $2.44\times10^{0}\pm6.50\times10^{-2}$\\
      282.07586 & $-2.30802$ & $3.13\pm1.06$ & $0.09\pm0.05$ & $0.56\pm0.06$ & $3830\pm84$      & $1.73\times10^{-3}\pm8.58\times10^{-4}$& $4.54\times10^{-9}\pm2.11\times10^{-9}$ & $3.57\times10^{0}\pm8.38\times10^{-1}$\\
      282.05382 & $-2.39705$ & $5.73\pm0.79$ & $0.58\pm0.77$ & $0.74\pm0.90$ & $4058\pm450$     & $2.43\times10^{-6}\pm1.21\times10^{-6}$& $1.42\times10^{-12}\pm7.06\times10^{-13}$& $2.82\times10^{0}\pm2.88\times10^{0}$\\
      282.05295 & $-2.35503$ & $1.77\pm1.47$ & $0.45\pm0.24$ & $1.94\pm0.89$ & $4589\pm832$     & $1.63\times10^{-3}\pm1.59\times10^{-2}$& $3.79\times10^{-9}\pm8.70\times10^{-8}$ & $9.96\times10^{0}\pm4.60\times10^{0}$\\
      281.99951 & $-2.28976$ & $14.18\pm0.58$& $0.18\pm0.21$ & $2.13\pm0.14$ & $4434\pm86$      & $6.68\times10^{-4}\pm3.12\times10^{-4}$& $1.70\times10^{-9}\pm8.47\times10^{-10}$& $2.78\times10^{1}\pm1.03\times10^{1}$\\
      \bottomrule
    \end{tabular}%
  }
    \vspace{2mm}
  \begin{minipage}{\textwidth}
    \footnotesize
    \textbf{Notes.}
Some uncertainties are reported as zero because, for those sources, no additional models satisfy the criterion $\chi^{2} - \chi^{2}_{\min} < 3$.

  \end{minipage}
\end{table*}

\subsection{Mass Spectrum}

In addition to the spectral energy distribution (SED) analysis in \ref{subsection:SEDS_of_YSOs}, we employ the $J$/$(J-H)$ colour–magnitude diagram (CMD) to investigate the mass distribution of young stellar objects (YSOs). CMDs incorporating the $K_{\mathrm{s}}$ band magnitude are deliberately excluded, as the $K_{\mathrm{s}}$-band is susceptible to significant contamination from near-infrared (NIR) excess emission arising from circumstellar material. This excess can lead to an overestimation of the intrinsic brightness of the sources, thereby introducing systematic errors in mass estimation.

\begin{figure}[htbp!]
    \centering
    \includegraphics[width=0.9\hsize]{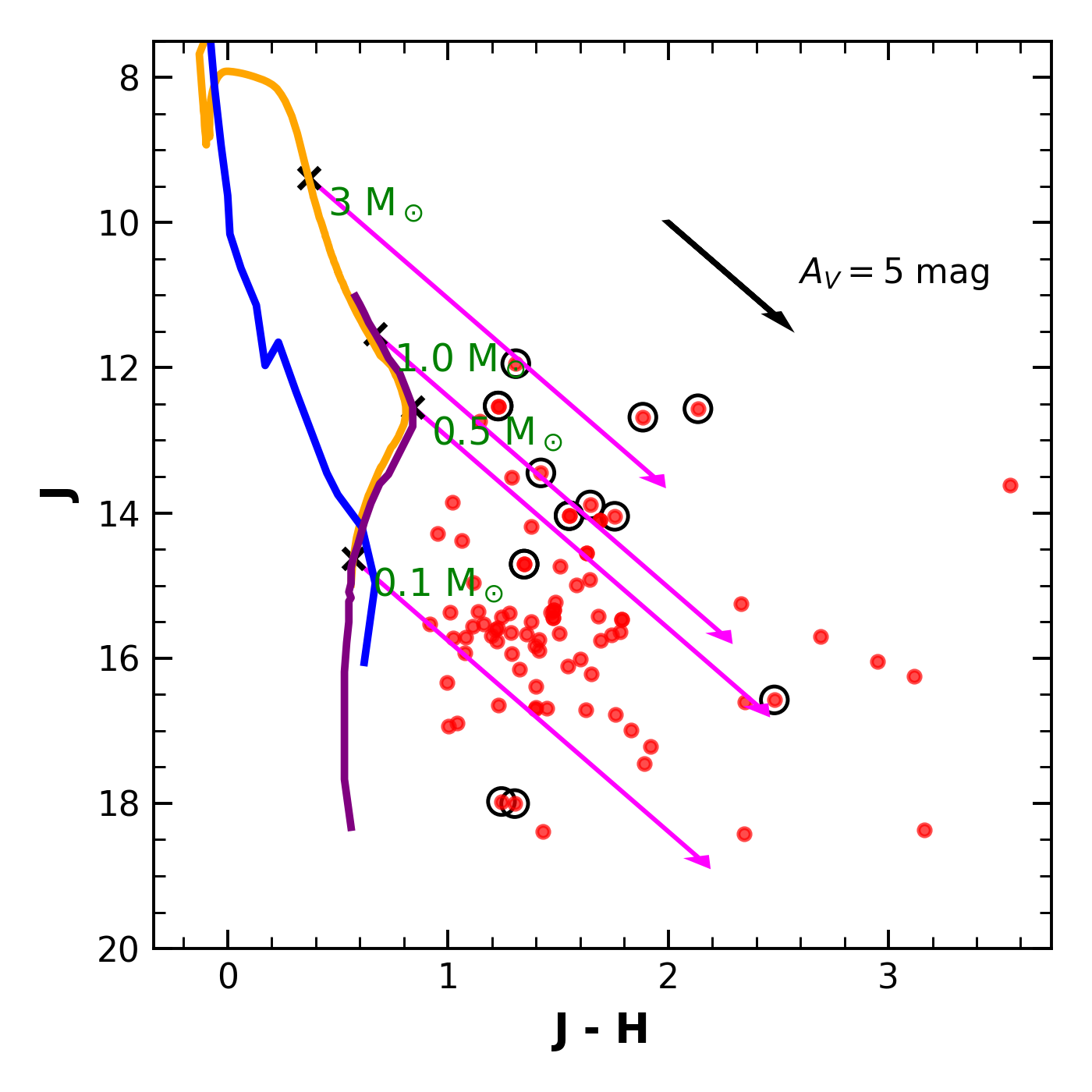}

    \caption*{(a) $J/(J - H)$ Colour--Magnitude Diagram showing the positions of the YSOs. The 1~Myr PMS isochrone from \citep{2015A&A...577A..42B} is shown in purple, the main-sequence dwarf locus by the blue thick line \citep{2000asqu.book.....C}, and the 1~Myr isochrone from \citep{10.1111/j.1365-2966.2012.21948.x} in orange. Black concentric circles mark YSOs with SED fitting.}

    \includegraphics[width=0.9\hsize]{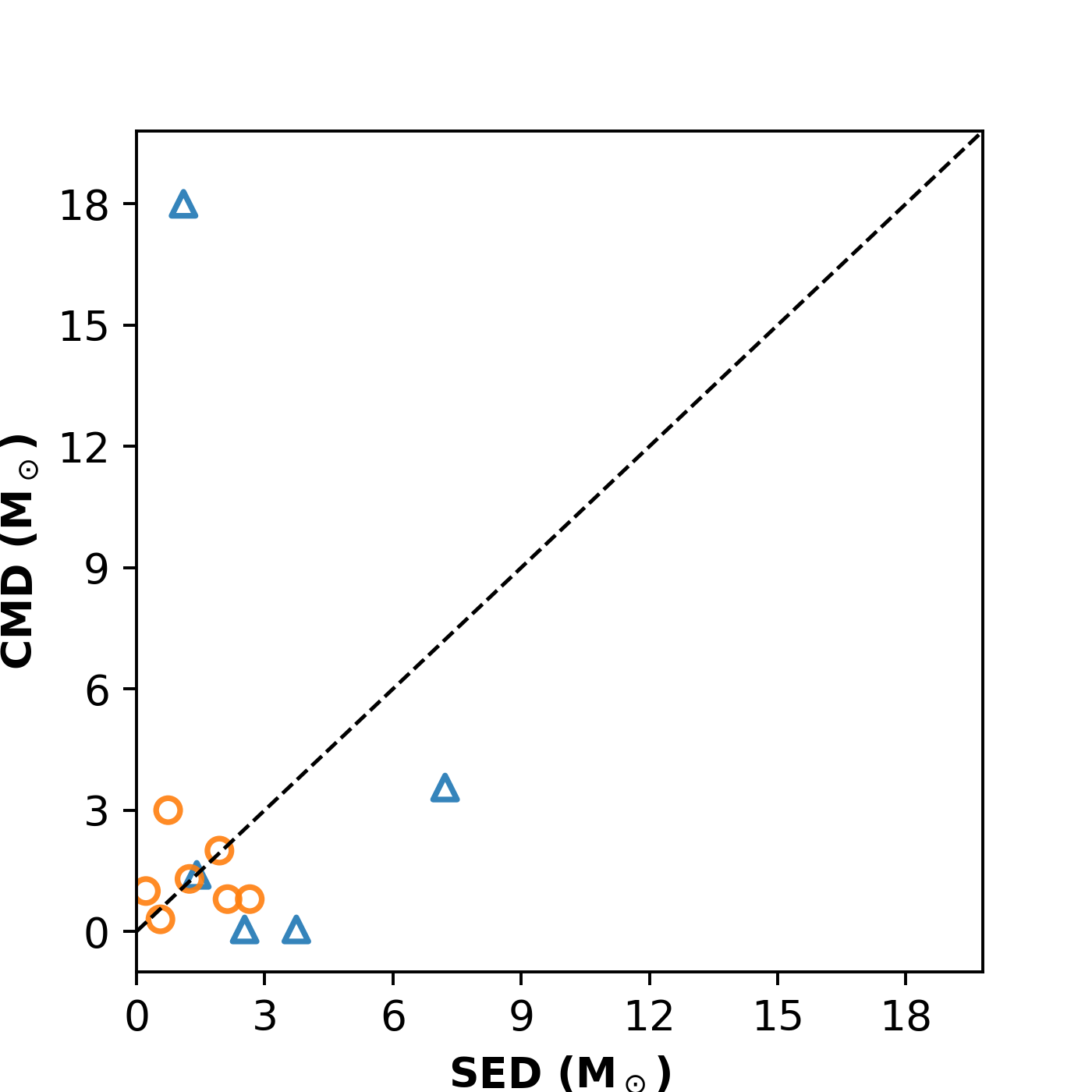}
    \caption*{(b) Comparison of the age estimates obtained from the CMD analysis with those from the SED fitting. Blue open triangles and orange open circles represent Class I and Class II sources, respectively.}

    \caption{Top: CMD of YSOs. Bottom: Correlation plot.}
    \label{fig:cmd_sed}
\end{figure}

Fig.~\ref{fig:cmd_sed}(a) shows the $J/(J-H)$ colour–magnitude diagram for all YSOs detected in both the $J$ and $H$ bands. Overlaid are two 1 Myr pre–main‐sequence isochrones: that of \citet{2015A&A...577A..42B}, which covers masses from $0.01$ to $1.4,M_\odot$, and that of \citet{10.1111/j.1365-2966.2012.21948.x}, extending from $0.1$ to $17.6,M_\odot$. Reddening vectors corresponding to PMS masses of $0.1$, $0.5$, $1.0$, and $3.0,M_\odot$ are also plotted. Objects for which SED fitting was carried out are highlighted with black circles.

As evident from the diagram, all but two of the SED-fitted YSOs fall within the mass range $\gtrsim 0.1\,M_\odot$. Broadly, the CMD suggests that most sources span a mass range of approximately $0.1$--$5\,M_\odot$. A correlation plot between masses derived from SED and CMD for corresponding 12 YSOs is presented in Fig.~\ref{fig:cmd_sed}(b).


The observed YSO colour spread likely arises from a mix of differential extinction and intrinsic age differences. Major uncertainties include the adopted distance ($\sim$606~pc), which scales absolute-to‐apparent magnitudes, unresolved binaries that can inflate luminosities, and systematic offsets in PMS evolutionary tracks that propagate into the derived masses.

    


   
\section{SPATIAL DISTRIBUTION OF YSO AND PROPERTIES OF THE CLOUD}
\label{sec:spatial distribution}
The intricate observational features seen in molecular clouds, such as filaments, bubbles, and irregular clumps, are believed to arise from various fragmentation processes. Star formation typically occurs within the dense cores of these clouds, and young stellar objects (YSOs) often trace the clumpy substructures of their parent environments \citep[e.g.,][]{1988ApJ...328..143L,1993AJ....105.1927G,2009ApJS..184...18G}. Their spatial distribution can be quantified by measuring typical separations and comparing them with the Jeans length for thermal fragmentation in a self-gravitating medium \citep{1993AJ....105.1927G}.
We extracted the overdense cores in our target regions to isolate local enhancements in surface density from the overall point-source distribution. Below, we outline the two methods used to analyze the spatial distribution of YSOs.  
\begin{figure}[htbp!]
    \centering

     \vspace{0.5cm}
    \includegraphics[width=0.9\hsize]{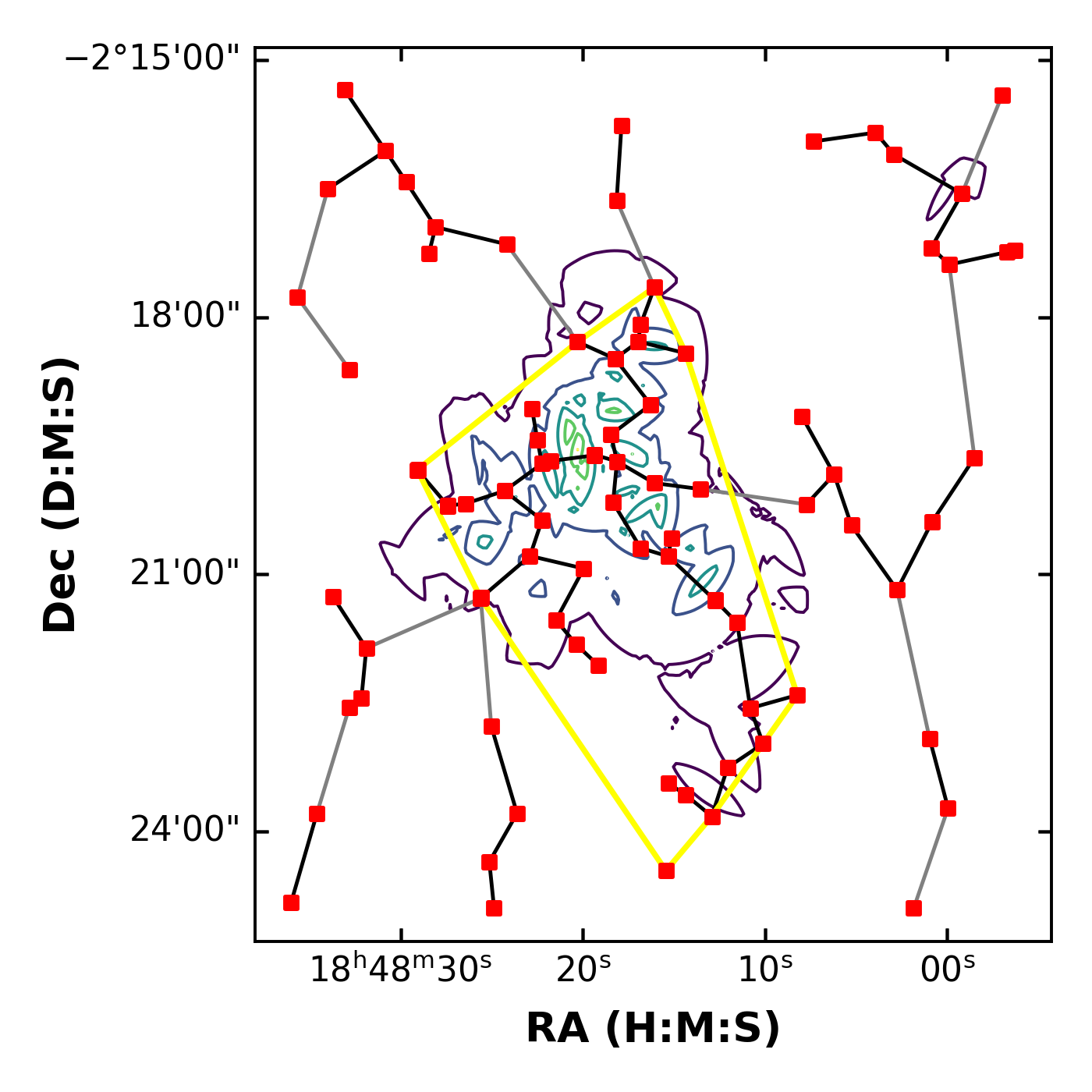}
    \caption*{(a) Iso-density contours (NN7) along with the Minimum Spanning Tree (MST) of the identified YSOs. Gray branches represent segments shorter than the critical length within the cores and the active region. The identified cores and the active region are enclosed by a solid yellow convex hull.}

 \vspace{1.2cm}
       
    \includegraphics[width=0.9\hsize, trim=0cm 0.4cm 0cm 0cm, clip]{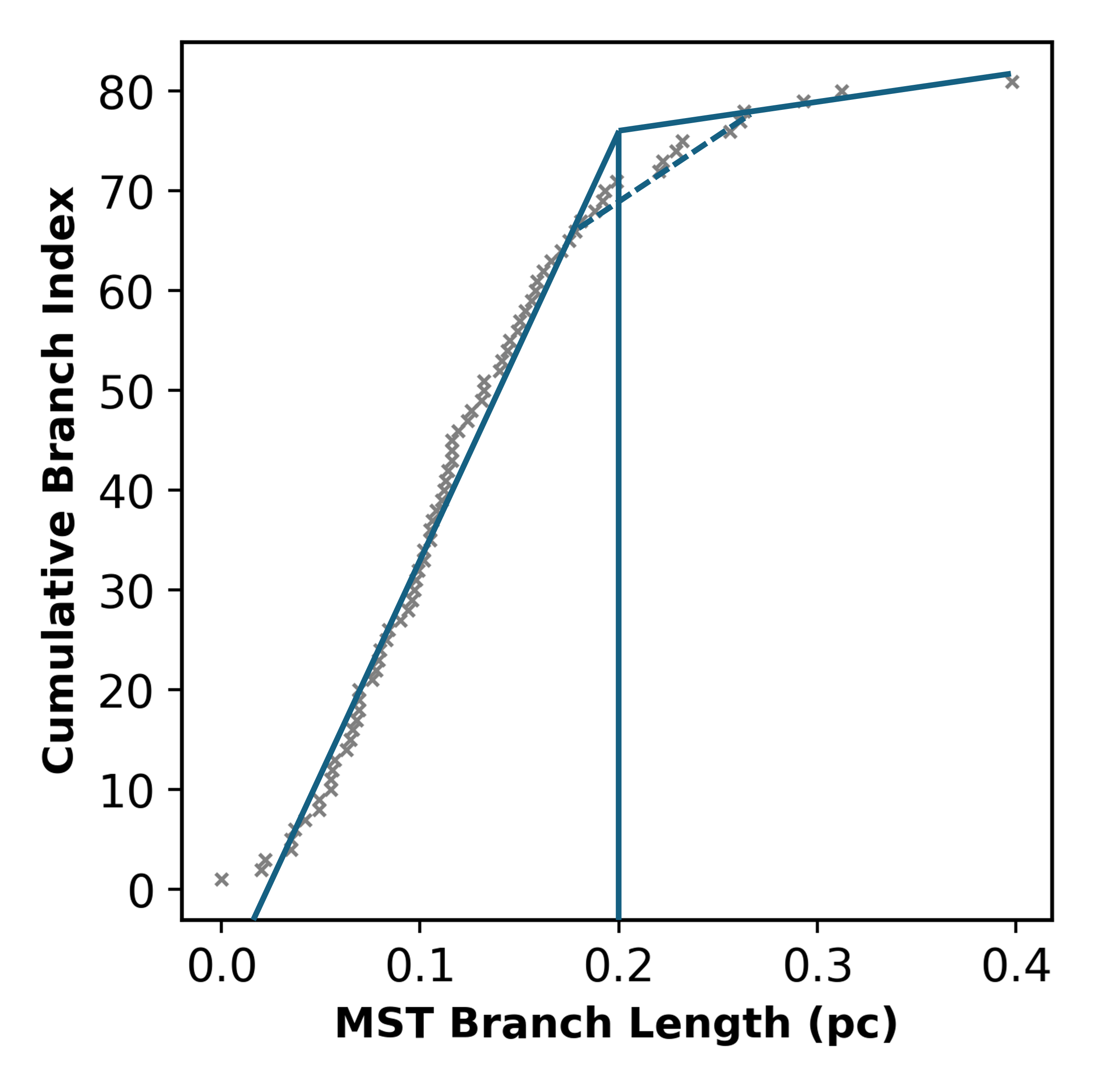}
    \caption*{(b) MST branch length distribution shown through a CDF plot. The straight lines represent linear fits to the points smaller and larger than the chosen critical branch length.}

   \caption{Top: MST branch length distribution. Bottom: Cumulative Distribution Function curve for YSOs.}

    \label{fig:mst_combined}
\end{figure}
   

   

\subsection{YSO Minimum Spanning Tree, Nearest Neighbor Distances, and Cluster Analysis}
The MST is defined as the network of lines, or branches, that
connect a set of points such that the total length of the branches
is minimized, and there are no closed loops \citep{2009ApJS..184...18G}. The MST branch
lengths represent the connections between sources (this is the
angular separation between two stars), and they can be analyzed
to identify substructures and groupings within the distribution Fig.~\ref{fig:mst_combined}(a).
By adopting a surface density threshold expressed as a critical
branch length, we can isolate cohesive structures (subclusters) within the star-forming region \citep[e.g.,][]{2004MNRAS.348..589C,2006A&A...449..151S, 10.1111/j.1365-2966.2007.12064.x}. To obtain a proper threshold distance, the two true line fit in the shallow and steep segment of the cumulative distribution function (CDF) for the branch length of the MST for YSOs is used (see Fig.~\ref{fig:mst_combined}(b)). The intersection point between these two lines is taken as the MST critical branch length \citep{2009ApJS..184...18G}. The regions which have a higher MST branch length are called as the active region (AR) where recent star formation took place or, which contains YSOs which moved out from the cores due to dynamical evolution. The critical branch length is defined as the point where the two lines in the CDF fit intersect. In this study, it was found to be around 0.2 pc (68 arcsecond).

To determine the stellar density, the coordinates of the stars were scatter-plotted and analyzed using the nearest neighbor method \citep{Gutermuth_2005}. The observed region was divided into a two-dimensional grid with a cell size of $3\arcsec$ \citep{Gutermuth_2005}, while the stellar surface density was estimated using the 7th nearest-neighbor method, whose effective resolution is independent of the grid size. For each position in the uniform grid, the radial distance required to encompass the $n$ nearest young stellar objects (YSOs) was calculated. The local surface density at each grid point was then estimated using the formula $\sigma = \frac{n - 1}{\pi r_n^2}$, with a fractional uncertainty of $(n - 2)^{-0.5}$ \citep{1985ApJ...298...80C}, where $\sigma$ represents the local stellar surface density and $r_n$ is the distance to the $n$-th nearest YSO from the grid point. In this study, $n = 7$ was adopted (see Fig.~\ref{fig:mst_combined}(a)) F; a larger $n$ provides better statistical sensitivity, while a smaller $n$ offers improved spatial resolution.

\begin{figure*}[h!]
  \centering
  \begin{minipage}[t]{0.64\textwidth}
    \vspace{0pt}
    \includegraphics[width=\linewidth]{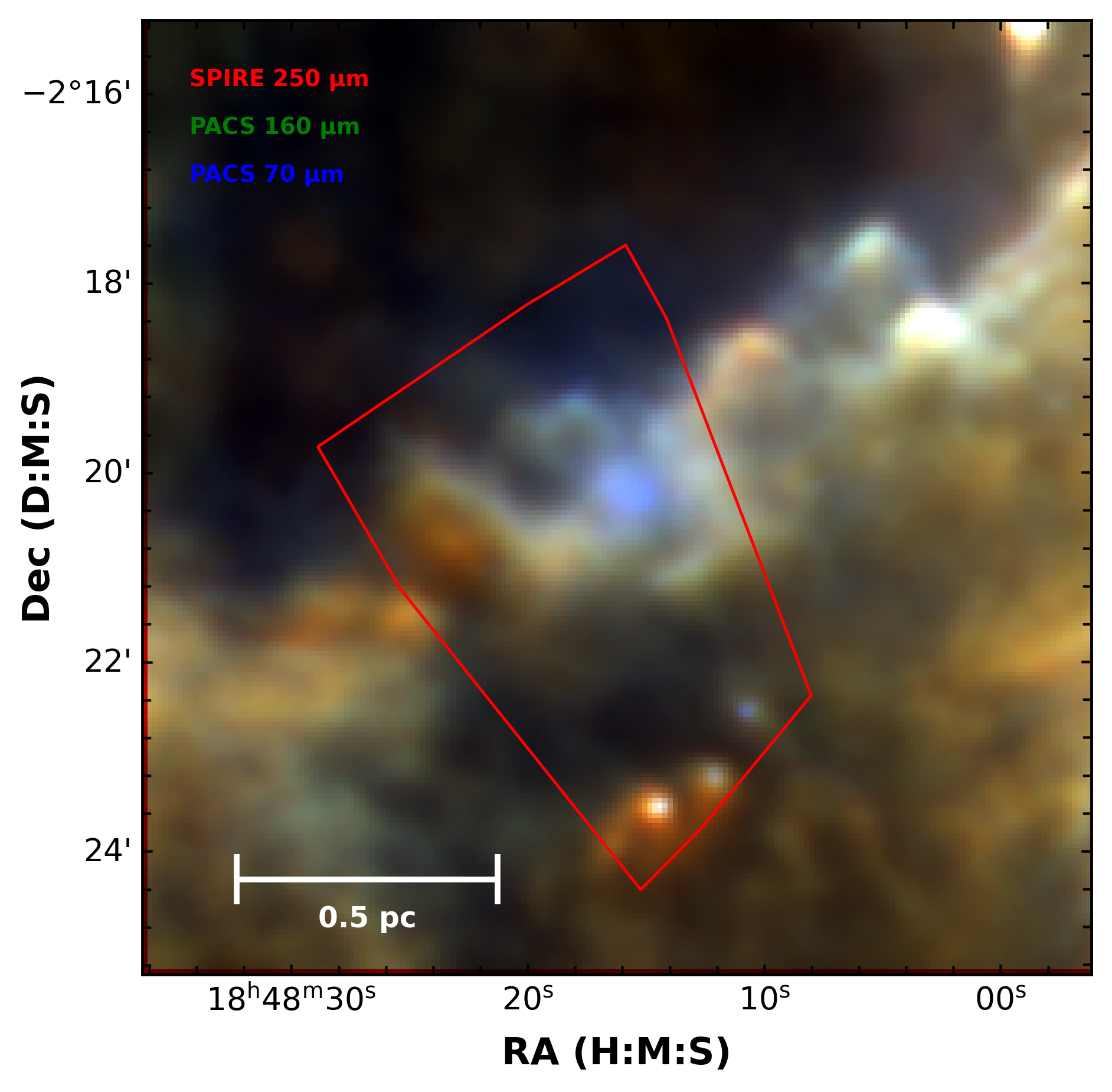}
  \end{minipage}
  \begin{minipage}[t]{0.32\textwidth}
    \vspace{0pt}
    \includegraphics[width=0.9\linewidth, trim=1.4cm 1.2cm 0cm 0cm, clip]{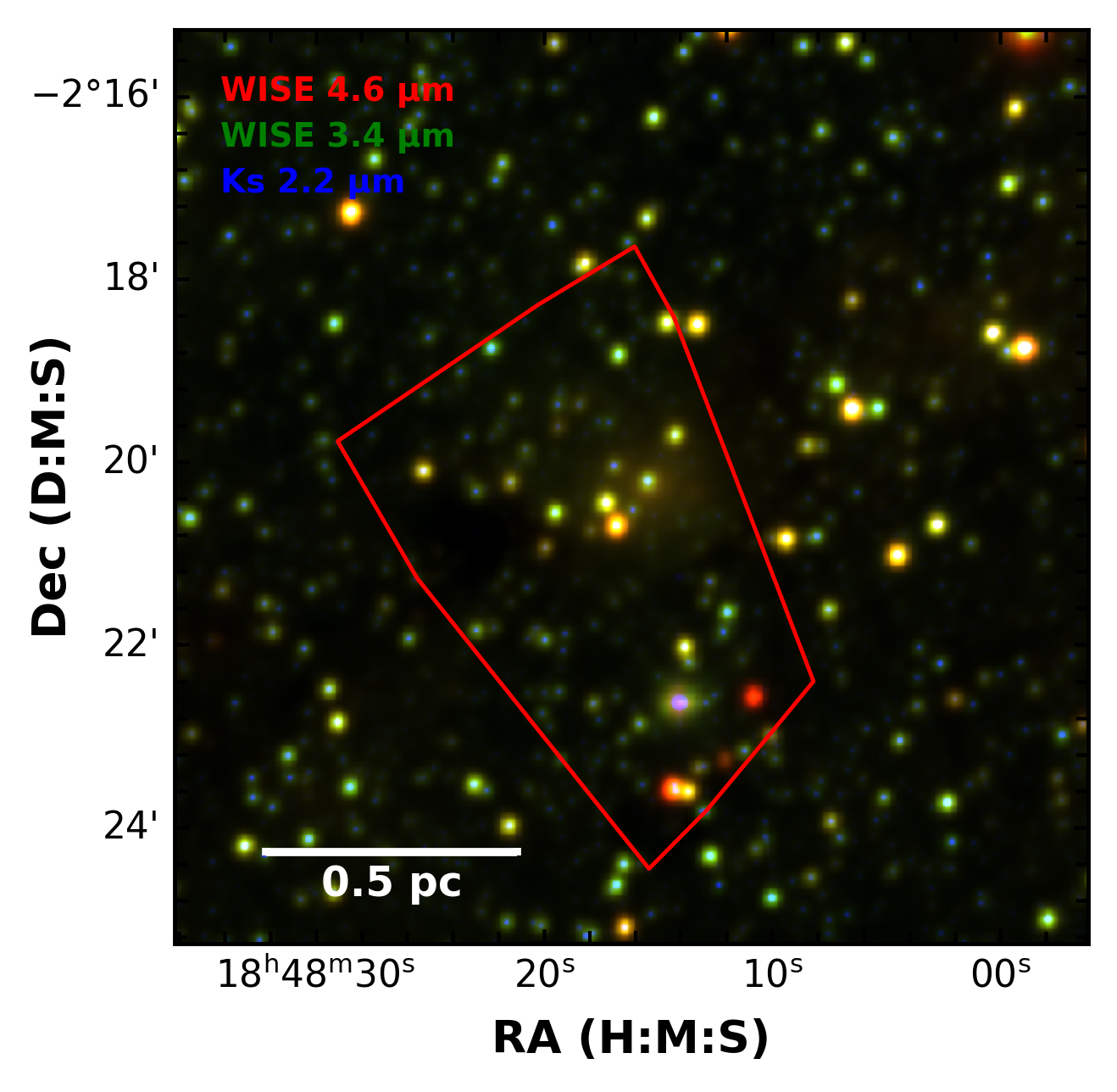}

    \includegraphics[width=0.9\linewidth, trim=1.4cm 1.2cm 0cm 0cm, clip]{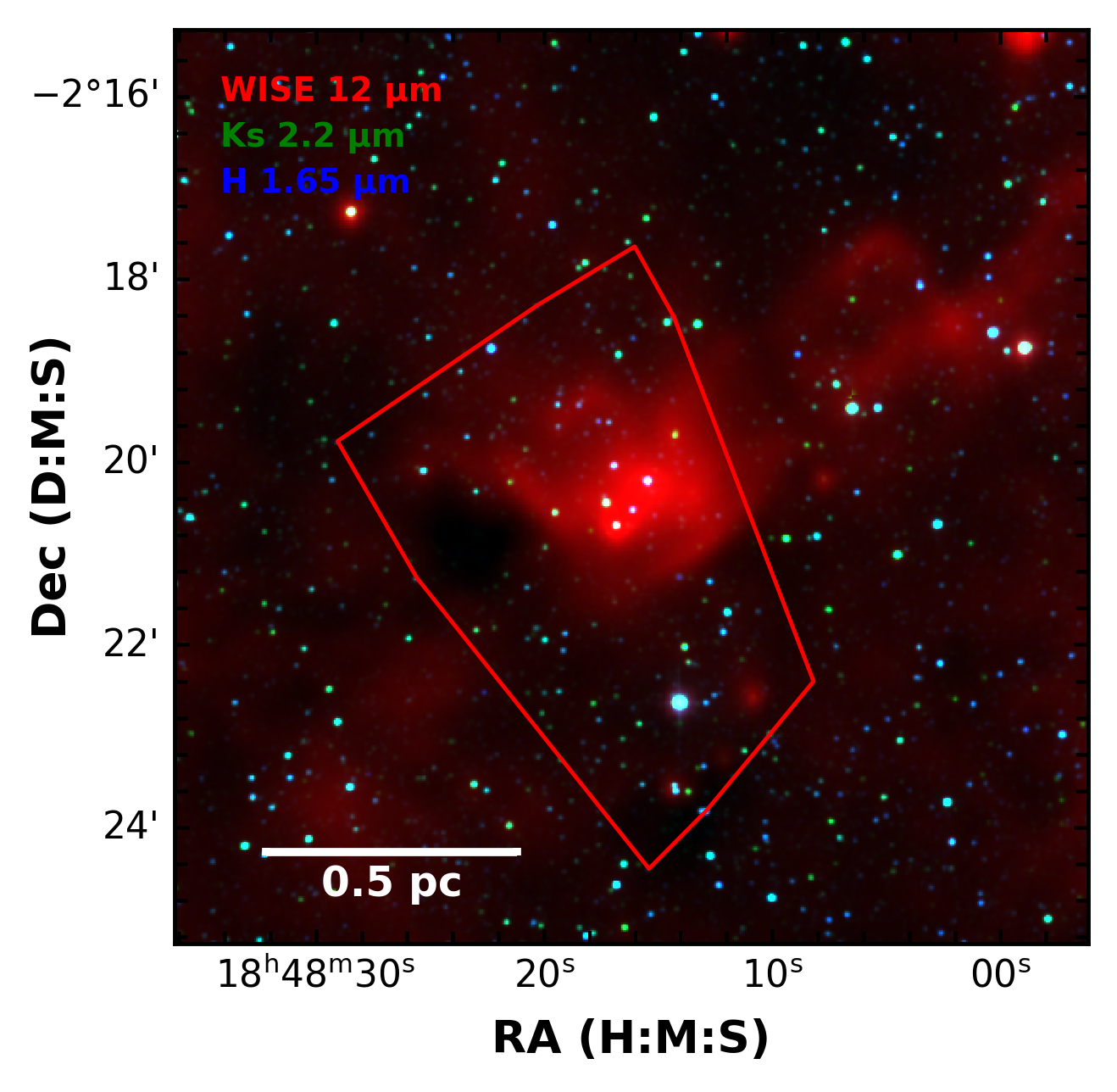}
  \end{minipage}

 \caption{
        (a) \textbf{Left panel:} \textit{Herschel} FIR colour-composite image covering an area of $\sim$10$\times$10 arcmin$^2$ centered on the cluster. The red polygon represents the convex hull outlining the cluster boundary based on the MST analysis.
        (b) \textbf{Right top panel:} Near-infrared colour-composite image created using the Ks (blue), \textit{WISE} 3.4~$\mu$m (green), and 4.6~$\mu$m  (red) bands. 
        (c) \textbf{Right bottom panel:} colour-composite image constructed using the H (blue), Ks (green), and \textit{WISE} 4.6~$\mu$m (red) bands. The red polygon denotes the same convex hull shown in the panel (b) and (a) for comparison.
    }
  \label{fig:aligned}
\end{figure*}

An apparent clustering of stars having a slightly elliptical geometry is visible, peaking at the central part of the cluster. We identified the overdense cores within each observed star‐forming region to isolate local peaks in surface density from the general point‐source distribution, and we find an inverse correlation between stellar positions and dust emission; in our analysis we quantify several key properties of the clusters and groups, namely their radial extent, mean stellar density, and surface density, to establish what is “typical” for young stellar cluster, where the circular radius \(R_{\mathrm{circ}}=0.619\,\mathrm{pc}\) is defined as half of the maximum separation between any two members (i.e., the radius of the minimal enclosing circle), the effective radius (spatial extent of the cluster) \(R_{\mathrm{hull}}=0.49\,\mathrm{pc}\) is computed via \(R_{\mathrm{hull}}=\sqrt{A/\pi}\) using the area \(A\) of the convex hull adjusted for the fraction of sources on its perimeter by \(A_{\mathrm{adjusted}}=A_{\mathrm{hull}}/(1-n_{\mathrm{hull}}/n_{\mathrm{total}})\) to account for elongated geometries, the aspect ratio of the distribution is given by \(R_{\mathrm{circ}}^2/R_{\mathrm{hull}}^2=1.567\), and the mean surface density \(\sigma_{\mathrm{mean}}=63.8\) $\mathrm{pc}^{-2}$ is obtained by dividing the number of \textit{Spitzer}‐identified IR-excess sources it encloses by adjusted convex-hull area \citep[see for details,][]{10.1093/mnras/staa2412}. Multi-wavelength composite images of the cluster region are presented in Fig.~\ref {fig:aligned}.

\subsection{NICER Method and Extinction Mapping}

Employing the NICER method from \citep{2001A&A...377.1023L} we derive extinction maps for our region using Near-Infrared photometric bands. NICER implements multiple colour measurements to reduce errors in the final extinction measurement. Our study uses the \textit{UKIDSS} $JHK$ data to construct the covariance matrix for the intrinsic colours of the combinations $J-H$ and $H-K$. The reddening vectors, $A_\lambda/A_V$, for the different colour combinations are taken from \citet{1981ApJ...249..481C}. The reddening parameter characterizes this variation \[
R_V = \frac{A_V}{E_{B-V}}, \]
 We adopt $R_V=3.1$ for all of the coefficients used in the NICER equations. Knowing the reddening vectors and intrinsic colours allows us to calculate the $A_V$ along the line of sight toward each background star.

\subsubsection{Intrinsic colours and Map Smoothing}
In previous studies \citep[e.g.,][]{1994ApJ...429..694L,2001A&A...377.1023L,2006A&A...454..781L,2009ApJ...703...52L}, intrinsic colours are usually approximated by averaging the observed colour of background stars from a nearby low-extinction region (also known as a control field). This approach also accounts for reddening due to intervening interstellar medium (ISM) dust. Nonetheless, selecting a low-extinction region is complicated because it covers the main structure of the cloud. We use a two-step approach to simulate a control field described in \citep{Li_2024, Chu_2021}.

Using Galactic simulations of stars from the TRILEGAL model\footnote{\url{http://stev.oapd.inaf.it/cgi-bin/trilegal}} \citep{2005A&A...436..895G}, we estimate the intrinsic colours of background stars in the direction of our region.

The input requires the equatorial or Galactic coordinates for the region of interest and a total field area of \(10~\arcmin \times 10~\arcmin\). A photometric system is chosen with a set magnitude limit, here a limiting magnitude of 19 in the $K$ band. Other parameters—such as the initial mass function, binary fraction, and Galactic components—were kept at their default values, and dust extinction was set to zero. The output is a list of stars with expected stellar parameters (e.g., temperature, mass) and predicted magnitudes in the \textit{2MASS} $JHK_{\mathrm{s}}$. The resulting distribution allows us to identify the expected intrinsic colours in $J-H$ and $H-K$.

The simulated intrinsic colours produced by the TRILEGAL model were inaccurate because the theoretical values do not account for interstellar dust along the line of sight to or from the molecular core. Therefore, we used the Galactic Dust Reddening and Extinction website \footnote{\url{ https://irsa.ipac.caltech.edu/applications/DUST/}}, which provides visual extinction estimates for a given location using data from the IRAS mission and the DIRBE experiment onboard the COBE satellite, to correct for this effect \citep{1998ApJ...500..525S}. We then summed the intrinsic and ISM colours to determine the total extinction. Errors in the intrinsic colours were taken as the standard deviation of the simulated values and propagated to estimate the uncertainty in the ISM extinction.

\begin{figure}[htbp!]
    \centering%
\hspace{-0.35cm}                \includegraphics[width=0.9\hsize]{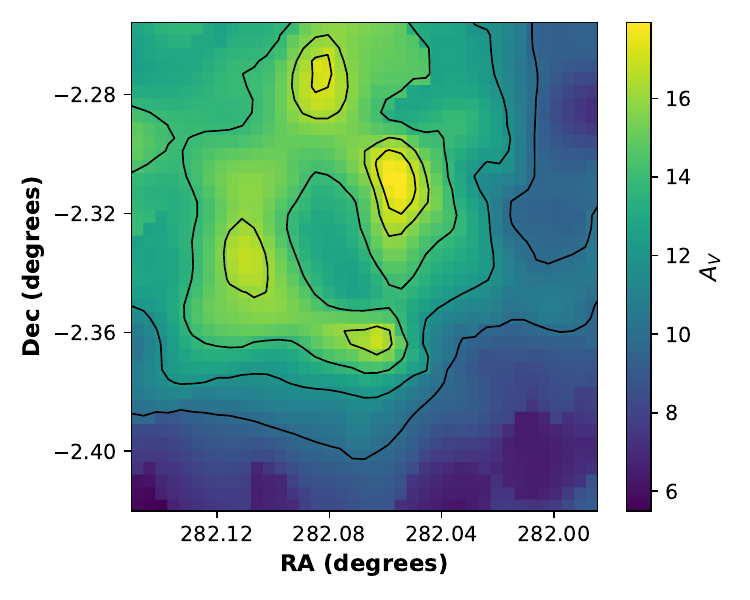}
    \caption{Extinction map obtained from the NICER technique, showing four extinction peaks; the northern peak likely arises from field‐star contamination, as it lacks a counterpart in the column‐density map.}
    \label{Fig10}%
\end{figure}

Once the extinction along each line of sight through the region has been calculated, we construct the extinction map (Fig.~\ref{Fig10}) using a $40 \times 40$ grid with a scale of $0.25~\arcmin$ per pixel and apply weighted‐mean smoothing with a Gaussian kernel characterized by a full width at half maximum (FWHM) of $0.5~\arcmin$. This method assumes that all stars are background objects, an assumption that is appropriate and does not introduce significant bias.

\subsection{\textit{Herschel} Column Density and Temperature Maps}

Thermal emission from cold dust in the far-infrared (FIR) wavelength range allows us to derive key physical parameters, such as dust temperature and column density, through spectral energy distribution (SED) modeling of thermal dust emission \citep[see for more detail,][]{2011A&A...535A.128B,2013A&A...551A..98L,2015MNRAS.447.2307M}. This analysis uses \textit{Herschel} FIR bands covering 160–500~$\mu$m (i.e., the Rayleigh–Jeans regime) to model the dust emission following these steps.

First, the surface brightness of all images was converted to Jy~pixel$^{-1}$. Since the PACS image was already in these units, only the SPIRE images (originally in Jy~beam$^{-1}$) required conversion, which was performed using the respective pixel scales. Next, the 160–350~$\mu$m images were convolved to the 500~$\mu$m image resolution (approximately 36~arcsec) using the convolution kernels of \citet{2011PASP..123.1218A} and regridded to a common 500 $\mu$m image. Note that the 70~$\mu$m image was not considered due to potential deviations from the optically thin assumption \citep{2014A&A...566A..45L}. The SED modeling employs modified blackbody fitting on a pixel-by-pixel basis. The observed flux density, is modeled as \citep{2011A&A...535A.128B,2012A&A...540A..10S,2012A&A...547A..11N,2013A&A...551A..98L}
\begin{equation}
S_\nu(\nu) - I_{\text{bg}}(\nu) = B_\nu(\nu, T_d) \left[ 1 - \exp\left(-\tau(\nu)\right) \right],
\label{eq:rad_transfer}
\end{equation}
where the optical length is given by,
\begin{equation}
\tau(\nu) = \mu_{\text{H}_2}\, m_{\text{H}}\, \kappa_\nu\, N(\text{H}_2),
\label{eq:optical_depth}
\end{equation}
Here, $\nu$ is the frequency, $B_\nu(\nu, T_d)$ is the Planck function at dust temperature $T_d$, and the solid angle corresponds to that subtended by a \(14~\arcsec \times 14~\arcsec\)pixel. The parameters $\mu_{\text{H}_2}$ (adopted as 2.8) and $m_{\text{H}}$ (the mass of hydrogen) are combined with the dust opacity, $\kappa_\nu$, and the column density, $N(\text{H}_2)$. The dust opacity is assumed to follow the relation
\begin{equation}
\kappa_\nu = 0.1\left(\frac{\nu}{1000\,\text{GHz}}\right)^{\beta} \, \text{cm}^2\,\text{g}^{-1}.
\label{eq:kappa}
\end{equation}
with $\beta = 2$ \citep{1983QJRAS..24..267H,1990AJ.....99..924B,2010A&A...518L.102A}. The model was fitted for each pixel using the four data points available, treating $T_d$ and $N(\text{H}_2)$ as free parameters. While higher-resolution column density maps can be obtained using Bayesian techniques such as PPMAP \citep{10.1093/mnras/stv2248}, which account for line-of-sight temperature variations, these methods are computationally intensive. For this study, we therefore adopt standard Herschel maps, which provide sufficient resolution to identify the relevant structures, while the use of PPMAP maps will be explored in future work.

\begin{figure}[htbp!]
   \centering
   \hspace{-0.75cm}
   \includegraphics[width=0.8\hsize]{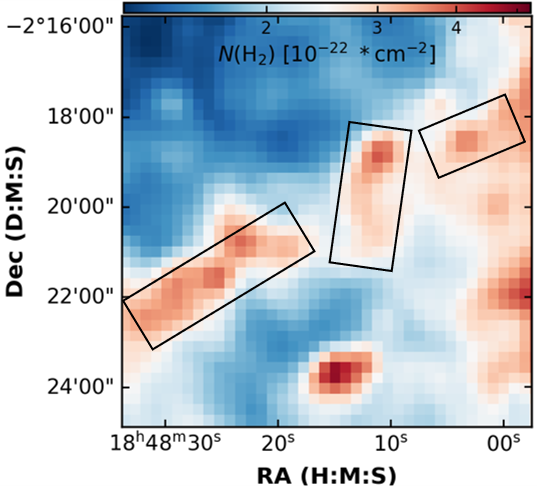}
      \caption{Column density map around
IRAS 18456-0223 derived from \textit{Herschel} images in colour
scale. Some pixels contained NaN values, likely due to bad or saturated pixels in the original \textit{Herschel} data. These were excluded from the analysis and masked in the final images to preserve data integrity. }
         \label{figcolumn}
   \end{figure}

\begin{figure}[h!]
   \centering
   \hspace{-0.75cm}
   \includegraphics[width=0.8\hsize]{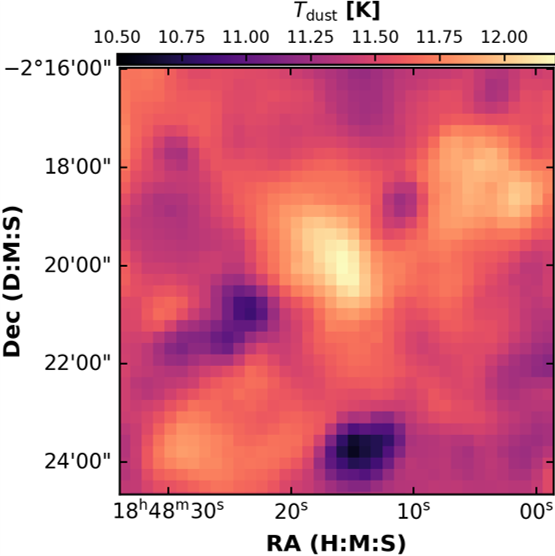}
      \caption{Dust temperature map}
         \label{figtemp}
   \end{figure}

The column‐density map (Fig.~\ref{figcolumn}) reveals four prominent peaks: a southern clump, the central IRAS 18456-0223 source, a south-western clump, and a northeastern clump. These peaks are interconnected by faint filamentary ridges of \(N(\mathrm H_2)\approx2\!-\!3\times10^{22}\,\mathrm cm^{-2}\). In contrast, as shown in Fig.~\ref{figtemp}, the temperature shows a distribution between 10.6 to 12.2~K, it is higher near the infrared cluster, peaking at $\sim$ 12.2 K.    

As expected, we see an overall similarity in the extinction peaks (Fig.~\ref{Fig10}) and areas of high column density (Fig.~\ref{figcolumn}). The northern
peak in the extinction map likely arises from field-star contamination, as it
lacks a counterpart in the column-density map. Detailed modeling is beyond the scope of this paper. 

\begin{figure}[h!]
   \centering
   
   \includegraphics[width=0.85\hsize]{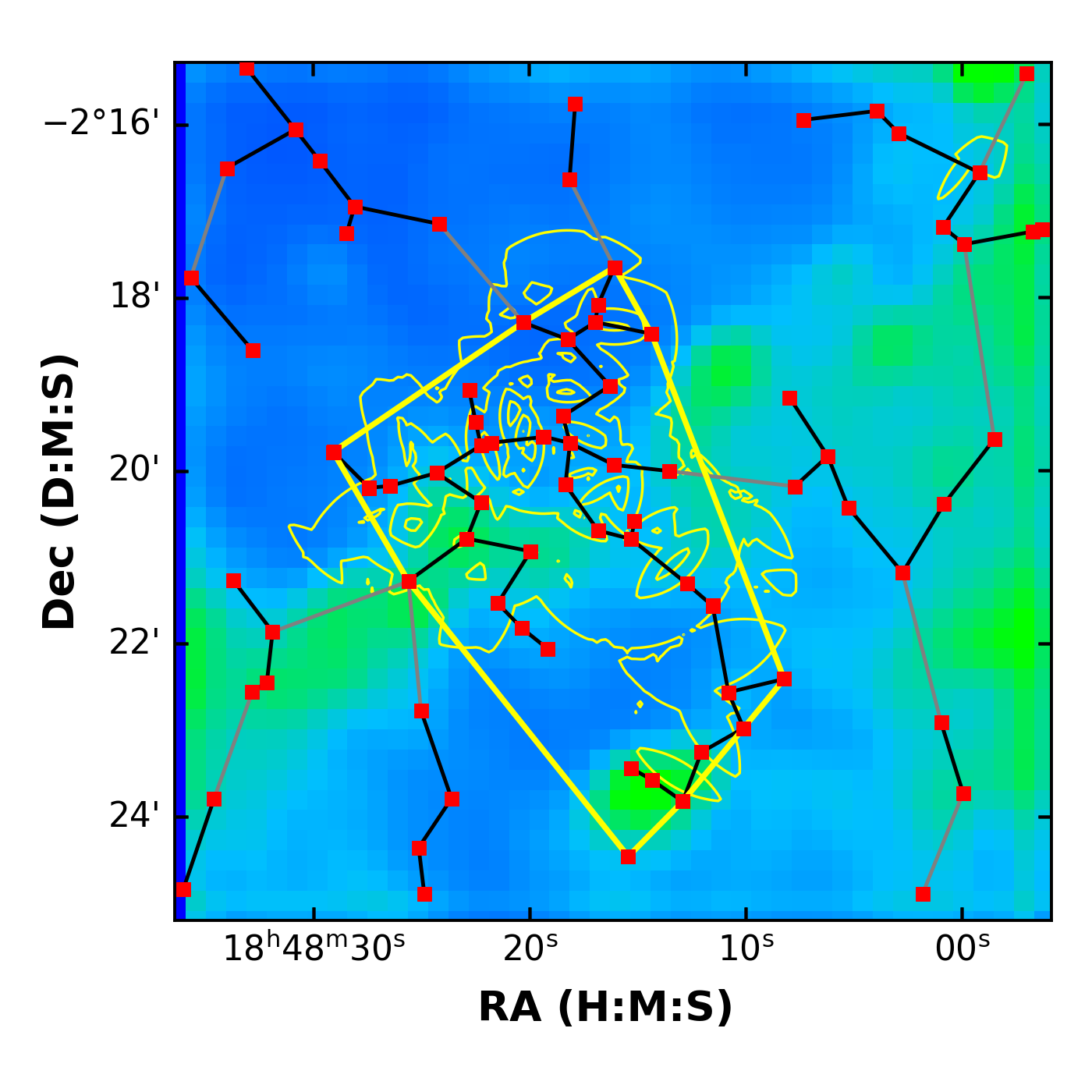}
      \caption{YSO distribution overlaid on the Herschel column density map, showing a tendency for YSOs to occupy lower-density regions.}
         \label{figmst_nh2}
   \end{figure}

Furthermore, when overlaying the YSOs distribution and MST structures on the Herschel column density map (Fig.~\ref{figmst_nh2}), we notice an anticorrelation between column density and YSO distribution, indicating that the YSOs tend to occupy comparatively lower-density regions, potentially due to the clearing of material by the ongoing star formation.

\vspace{1cm}


\section{OPTICAL SPECTRA \& DISCUSSION}
\label{sec:discussions}
\begin{figure}[htbp!]
   \centering
   
   \includegraphics[width=0.85\hsize]{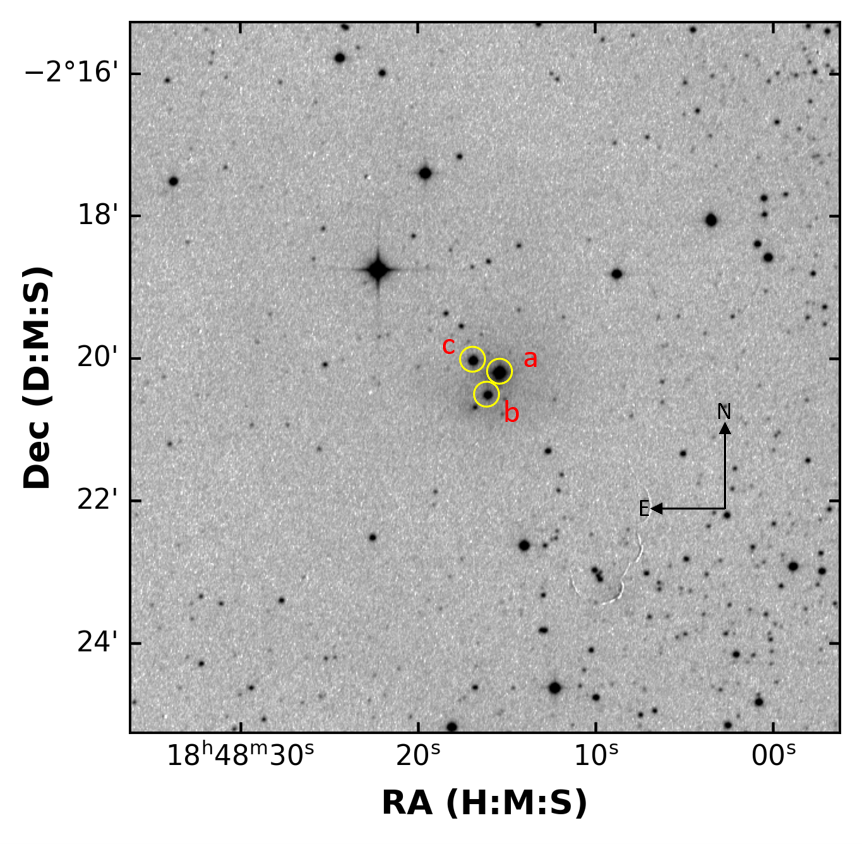}
      \caption{ POSS‑II UKSTU infrared image ($\lambda = 8500\,$\AA) of the target field with stars labelled \textbf{`a'}, \textbf{`b'}, and \textbf{`c'} in bold red. These three stars were specifically observed with optical spectroscopy as reference targets. }
         \label{fig3stars}
   \end{figure}

\begin{figure*}[h!]
    \centering
    \resizebox{18cm}{8cm}{%
    \includegraphics{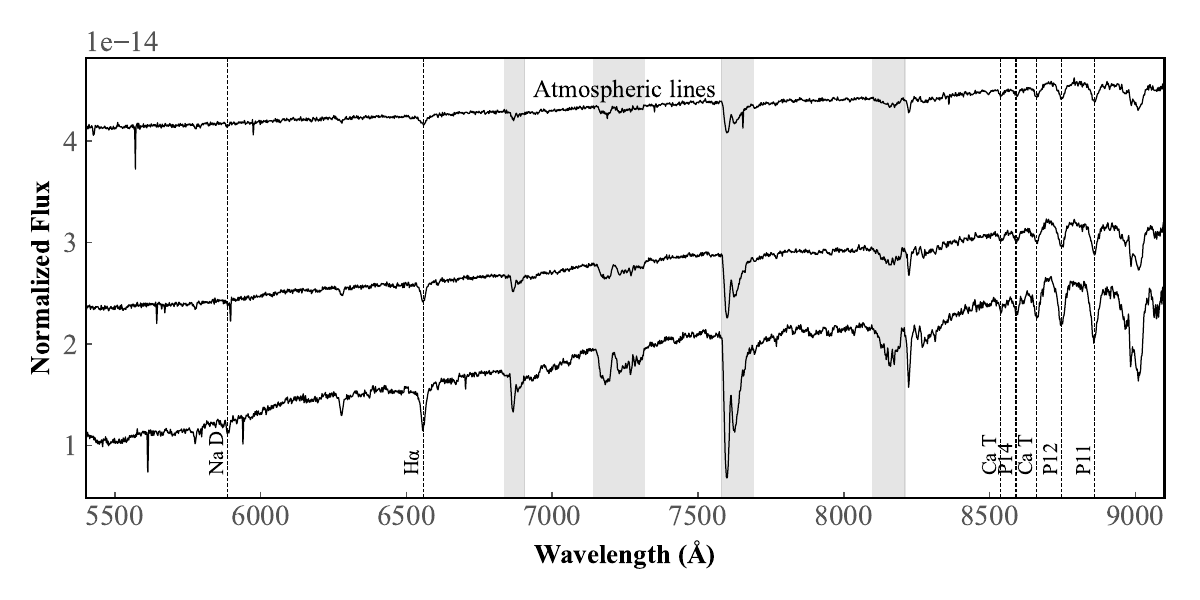}}
\caption{Optical spectra of the three bright sources located in the central region of the IRAS field. The sources labeled as \textbf{`a'} (bottom), \textbf{`b'} (middle), and \textbf{`c'} (top) correspond to the targets marked in red in Fig.~\ref{fig5ap}. Prominent spectral lines found in the analysis are indicated.}
      \label{fig5ap}
\end{figure*}

\begin{figure}[h!]
   \centering
   
   \includegraphics[width=0.9\hsize]{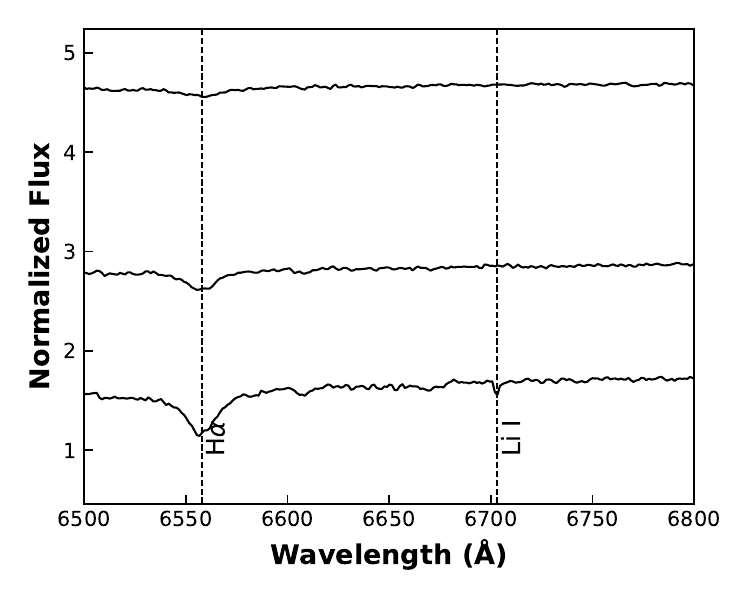}
      \caption{Optical spectra of sources `a' (bottom), `b' (middle), and `c' (top) over 6500–6800\,\AA, vertically offset for clarity. The Li\,\textsc{i} 6707\,\AA\ absorption line is prominently detected only in source `a'.}

         \label{fig3spectra}
   \end{figure}

 Using USNO BR and \textit{DENIS} $IJK$ magnitudes, \cite{2001A&A...370..991K} estimated the spectral types of the stars `a', `b', and `c' (see Fig.\ref{fig3stars}) as late-G to late-M III, K3-6 III and K3-6 III, respectively, with a foreground extinction of around 2 magnitudes in the $V$ band. Now, from the \textit{Gaia DR3} parallaxes, we have estimated the distances of these stars as 615, 632 and 627 pc, respectively. Then, using the \textit{2MASS} $K_{\mathrm{s}}$ magnitudes of these stars, we estimate their spectral types to be around B8 V, A1 V and A5 V, respectively. This calculation is valid for extinctions of 0-0.22 in the $K_{\mathrm{s}}$ band. 
However, stars `a' (SSTOERC G030.4898-00.3581) and `c' (SSTGLMC G030.4863-00.3629) are YSO candidates as per \citet{Saral_2015} and \citet{2021ApJS..254...33K}, respectively.  
However, they were not detected as YSOs as per the technique we followed in Sec.\ref{sec:YSO population}.
Our optical spectra of these sources can be seen in Fig.~\ref{fig5ap}. All of them show only absorption lines such as H$\alpha$, Ca~II triplet, Paschen-series and NaI D.  Emission lines that are  usually seen in accreting young stars are absent. In addition, in Fig.~\ref{fig3spectra} we identify the Li I 6707 Å  absorption feature , having an equivalent width of 0.2\,\AA , in source `a'. In such cases, the presence of this line is an indicator of stellar youth. In Gaia 17bpi,  this line had an equivalent width of 0.47\,\AA \citep{2018ApJ...869..146H}, and in the seven newly-discovered and six known members of the TW Hya association it ranges from 0.36 to 0.57\,\AA \citep{Webb_1999}. Therefore, the Li I line in star `a' is weaker in comparison. More observations are needed to monitor the possible temporal variation of the equivalent width of this line in star `a' and to see if the line appears in stars `b' and `c'.

Helium lines are not apparent in the spectra of stars a-c implying that the stars are of spectral type later than B. A visual inspection of the spectra indicate that the stars are of A-F spectral type. Then, we compared our spectra  with those in the templates
of stars from the stellar library described in \citet{2006MNRAS.371..703S} and provided by Miles Stellar Library\footnote{\url{https://research.iac.es/proyecto/miles/pages/stellar-libraries/miles-library.php}}. Before comparing, we
normalized the spectra of our sources and also used $A_V = 0-2$\, for dereddening to account for the uncertainty in extinction. The stars are found to be of K 4-5 III spectral type by visual comparison. But, giants at a distance of around 600 pc would have much brighter apparent magnitudes ; hence these stars cannot be giants. The optical spectra of FU Ori, V1057 Cyg and V1515 Cyg resembled F or G supergiants \citep{2008AJ....135..637H}. The FUor Gaia-17bpi displayed a GK-type spectrum in the optical and an M-type spectrum in the infrared during its outburst phase \citep{2018ApJ...869..146H}. The optical spectra of PTF 14jg resembled a reddened late F or early G supergiant, but also showed  early A- and B-type features, thus eluding a strict classification \citep{Hillenbrand_2019}. All these stars showed the Li 6707\,\AA\, line. 
 SED fitting for the sources a-c was performed following the method described in Sec.~\ref {subsection:SEDS_of_YSOs}, and the results, which indicate that they are young stars, are summarized in Table~\ref{tab:sed_errors_compact}. The exercise using optical spectra demonstrates the difficulty in assigning spectral types based on limited wavelength coverage. In contrast, as seen in Sec.\ref{subsection:SEDS_of_YSOs}, even a wide wavelength coverage has the limitation of non-contemporaneity.

As noted earlier, IRAS 18456-0223 lies in the general direction of the W43 massive star-forming region, which  
lies at a distance of 5.49 (trigonometric parallax; \citet{Zhang_2014}) -- 6 kpc (kinematic; \citet{2011A&A...529A..41N}). As per \cite{2001A&A...370..991K} the star-forming cloud consisting of their 28 \textit{DENIS} sources is situated at a distance of 1600$\pm$400 pc. Since none of these sources have Gaia counterparts, it is not possible to independently estimate their distances via the parallax method. The YSOs we have identified are at around 600 pc.   Therefore, it appears that there are two or three areas of star formation in that line of sight. Consequently, the results presented in Sec.\ref{sec:spatial distribution} must be seen in this context. A similar conundrum, caused by the mismatch between distance and spectral type modeled using the SED, was noticed in the case of the FUor 2MASS J06593158-0405277 by \citep{2015ApJ...801L...5K}. They concluded that this star was not associated with the LDN 1650 molecular cloud, which is at a distance of 2.3~kpc, but is located much closer at 450~pc.

The flaring star is undetected in Gaia and the infrared surveys we have considered, so its location and properties cannot be determined.

\begin{table*}
\centering
\small 
\setlength{\tabcolsep}{3pt} 
\caption{Properties of the three bright sources via SEDFITTING.}
\label{tab:sed_errors_compact}
\resizebox{\textwidth}{!}{
\begin{tabular}{cccccccccc}
\hline
ID & RA (deg) & Dec (deg) & $A_V$ (mag) & Age & Mass & $T_{\rm eff}$ (K) & $M_{\rm disk}$ & $\dot{M}_{\rm disk}$ & $L_{\rm tot}$ \\
\hline
a & 282.0645 & $-$2.3368 & 3.93$\pm$0.49 & 0.96$\pm$1.47 & 4.60$\pm$0.29 & 14410$\pm$1215 & 7.25e$-$07$\pm$2.59e$-$07 & 1.35e$-$13$\pm$9.51e$-$14 & 4.90e+02$\pm$22.6 \\
b & 282.0671 & $-$2.3420 & 3.92$\pm$1.69 & 0.37$\pm$0.19 & 4.25$\pm$1.19 & 4939$\pm$473 & 8.54e$-$05$\pm$2.98e$-$02 & 1.23e$-$09$\pm$3.57e$-$07 & 4.85e+01$\pm$29.3 \\
c & 282.0707 & $-$2.3341 & 1.00$\pm$3.67 & 0.52$\pm$1.86 & 3.32$\pm$0.70 & 4870$\pm$5545 & 7.72e$-$04$\pm$3.86e$-$04 & 6.77e$-$10$\pm$3.38e$-$10 & 2.10e+01$\pm$183 \\
\hline
\end{tabular}
}
\end{table*}

\section{Conclusions}
\label{sec:conclusions}
We studied a \(10~\arcmin \times 10~\arcmin\)  region around 
IRAS 18456-0223, using  archival
data from \textit{Spitzer}, \textit{WISE}, \textit{Herschel}, \textit{2MASS}, \textit{UKIDSS}, and \textit{Gaia DR3}, as well as our  optical spectroscopy data from \textit{HCT}.
\begin{enumerate}
\setlength\itemsep{0em}
\setlength\parskip{0em}
\setlength\parsep{0em}
\item {\it YSO census and properties:} We identify 89 young stellar objects via infrared excess, comprising 80 class II and 9 class I sources. This classification is based on standard IR colour criteria, separating protostars from disk-bearing pre-main-sequence stars. Gaia proper motions show that the IR-excess sources cluster in the velocity space, yielding a median distance $\sim606\,\mathrm{pc}$. Multiwavelength SED fitting yields a wide age range (upto $4$~Myr) and stellar masses $\sim0.1$–$7.2$~$M_\odot$, implying a population from very young protostars to T Tauri stars. We compare \textit{UKIDSS} and \textit{2MASS} data of the YSOs and find that some of them show variability, as is expected. The location and nature of the flaring star remain unknown.

\item {\it Spatial distribution:} The YSO population is concentrated with an effective radius of order $0.5\,\mathrm{pc}$ and mean surface densities $\sim60\,\mathrm{pc}^{-2}$, similar to those of known embedded clusters.

\item {\it Spectroscopic signatures:} Optical spectra of three bright sources near the flaring star show them to be A-K type. One of them shows the Li I 6707\,\AA\, line, indicating its youth. SED fitting indicates that all of them are young stars. 

\item {\it Cloud properties:} \textit{Herschel}-based dust maps reveal multiple column density peaks. The NIR extinction map exhibits several high-$A_V$ peaks. Only some of them coincide with the sub-mm column density maxima, tracing the same dense cores. This clumpy, filamentary morphology is characteristic of active star-forming clouds.
\item{\it General : } It appears that there are at least two unconnected star-forming regions in the line-of-sight to IRAS 18456-0223.
\end{enumerate}

\section*{Acknowledgements}
     This work makes use of archival data from the \textit{Spitzer} Space Telescope (GLIMPSE survey), the Wide-field Infrared Survey Explorer (\textit{WISE})—a joint project of the University of California, Los Angeles, and the Jet Propulsion Laboratory/California Institute of Technology, funded by NASA—and the \textit{Herschel} Space Observatory. We also utilised data from the Two Micron All Sky Survey (\textit{2MASS}), the UKIRT Infrared Deep Sky Survey (\textit{UKIDSS}), and \textit{Gaia DR3}, which have been essential for photometric and astrometric analysis. This research used the Python-based SED fitting tool developed by Robitaille, and accessed resources provided by SIMBAD, the VizieR catalogue, and the CDS cross-match service, operated at CDS, Strasbourg, France. The work was carried out at the Indian Institute of Astrophysics (IIA), and NP gratefully acknowledges IIA for providing financial support and computational resources. We thank the referees for their constructive suggestions which helped to improve the paper.

\bibliography{ref.bib}

@article{2001A&A...370..991K,
       author = {{Kimeswenger}, S. and {Weinberger}, R.},
        title = "{IRAS 18456-0223 - a flare star in a new star forming region}",
      journal = {\aap},
         year = 2001,
        month = may,
       volume = {370},
        pages = {991-995},
          doi = {10.1051/0004-6361:20010287},
       adsurl = {https://ui.adsabs.harvard.edu/abs/2001A&A...370..991K}
}

@ARTICLE{2007MNRAS.380.1141S,
       author = {{Sharma}, Saurabh and {Pandey}, A.~K. and {Ojha}, D.~K. and {Chen}, W.~P. and {Ghosh}, S.~K. and {Bhatt}, B.~C. and {Maheswar}, G. and {Sagar}, Ram},
        title = "{Star formation in young star cluster NGC1893}",
      journal = {\mnras},
         year = 2007,
        month = sep,
       volume = {380},
       number = {3},
        pages = {1141-1160},
          doi = {10.1111/j.1365-2966.2007.12156.x},
archivePrefix = {arXiv},
       eprint = {0707.0269},
 primaryClass = {astro-ph},
       adsurl = {https://ui.adsabs.harvard.edu/abs/2007MNRAS.380.1141S}
}

@misc{hillenbrand2002youngcircumstellardisksevolution,
      title={Young Circumstellar Disks and Their Evolution: A Review}, 
      author={Lynne A. Hillenbrand},
      year={2002},
      eprint={astro-ph/0210520},
      archivePrefix={arXiv},
      primaryClass={astro-ph},
      url={https://arxiv.org/abs/astro-ph/0210520}, 
}

@ARTICLE{2008MNRAS.383.1241P,
       author = {{Pandey}, A.~K. and {Sharma}, Saurabh and {Ogura}, K. and {Ojha}, D.~K. and {Chen}, W.~P. and {Bhatt}, B.~C. and {Ghosh}, S.~K.},
        title = "{Stellar contents and star formation in the young star cluster Be 59}",
      journal = {\mnras},
         year = 2008,
        month = jan,
       volume = {383},
       number = {3},
        pages = {1241-1258},
          doi = {10.1111/j.1365-2966.2007.12641.x},
archivePrefix = {arXiv},
       eprint = {0710.5429},
 primaryClass = {astro-ph},
       adsurl = {https://ui.adsabs.harvard.edu/abs/2008MNRAS.383.1241P}
}

@article{Ojha_2004a,
doi = {10.1086/420876},
url = {https://doi.org/10.1086/420876},
year = {2004},
month = {jun},
publisher = {},
volume = {608},
number = {2},
pages = {797},
author = {Ojha, D. K. and Tamura, M. and Nakajima, Y. and Fukagawa, M. and Sugitani, K. and Nagashima, C. and Nagayama, T. and Nagata, T. and Sato, S. and Pickles, A. J. and Ogura, K.},
title = {Deep Near-Infrared Observations of the W3 Main Star-forming Region},
journal = {The Astrophysical Journal}
}

@ARTICLE{1989ApJ...340..265W,
       author = {{Wood}, Douglas O.~S. and {Churchwell}, Ed},
        title = "{Massive Stars Embedded in Molecular Clouds: Their Population and Distribution in the Galaxy}",
      journal = {\apj},
         year = 1989,
        month = may,
       volume = {340},
        pages = {265},
          doi = {10.1086/167390},
       adsurl = {https://ui.adsabs.harvard.edu/abs/1989ApJ...340..265W}
}

@article{Hillenbrand_2019,
doi = {10.3847/1538-4357/ab06c8},
url = {https://doi.org/10.3847/1538-4357/ab06c8},
year = {2019},
month = {mar},
publisher = {The American Astronomical Society},
volume = {874},
number = {1},
pages = {82},
author = {Hillenbrand, Lynne A. and Miller, Adam A. and Carpenter, John M. and Kasliwal, Mansi M. and Isaacson, Howard and Tang, Sumin and Joshi, Vishal and Banerjee, D. P. K. and Cutri, Roc M.},
title = {PTF14jg: The Remarkable Outburst and Post-burst Evolution of a Previously Anonymous Galactic Star},
journal = {The Astrophysical Journal}
}

@article{Webb_1999,
doi = {10.1086/311856},
url = {https://doi.org/10.1086/311856},
year = {1999},
month = {feb},
publisher = {},
volume = {512},
number = {1},
pages = {L63},
author = {Webb, R. A. and Zuckerman, B. and Platais, I. and Patience, J. and White, R. J. and Schwartz, M. J. and McCarthy, C.},
title = {Discovery of Seven T Tauri Stars and a Brown Dwarf Candidatein the Nearby TW Hydrae Association},
journal = {The Astrophysical Journal}
}

@ARTICLE{2018ApJ...869..146H,
       author = {{Hillenbrand}, Lynne A. and {Contreras Pe{\~n}a}, Carlos and {Morrell}, Sam and {Naylor}, Tim and {Kuhn}, Michael A. and {Cutri}, Roc M. and {Rebull}, Luisa M. and {Hodgkin}, Simon and {Froebrich}, Dirk and {Mainzer}, Amy K.},
        title = "{Gaia 17bpi: An FU Ori-type Outburst}",
      journal = {\apj},
         year = 2018,
        month = dec,
       volume = {869},
       number = {2},
          eid = {146},
        pages = {146},
          doi = {10.3847/1538-4357/aaf414},
archivePrefix = {arXiv},
       eprint = {1812.06640},
 primaryClass = {astro-ph.SR},
       adsurl = {https://ui.adsabs.harvard.edu/abs/2018ApJ...869..146H}
}

@ARTICLE{2015ApJ...801L...5K,
       author = {{K{\'o}sp{\'a}l}, {\'A}. and {{\'A}brah{\'a}m}, P. and {Mo{\'o}r}, A. and {Haas}, M. and {Chini}, R. and {Hackstein}, M.},
        title = "{The Progenitor of the FUor-Type Young Eruptive Star 2MASS J06593158-0405277}",
      journal = {\apjl},
         year = 2015,
        month = mar,
       volume = {801},
       number = {1},
          eid = {L5},
        pages = {L5},
          doi = {10.1088/2041-8205/801/1/L5},
archivePrefix = {arXiv},
       eprint = {1501.07735},
 primaryClass = {astro-ph.SR},
       adsurl = {https://ui.adsabs.harvard.edu/abs/2015ApJ...801L...5K}
}

@ARTICLE{1993ApJ...407..657C,
       author = {{Carpenter}, John M. and {Snell}, Ronald L. and {Schloerb}, F.~P. and {Skrutskie}, M.~F.},
        title = "{Embedded Star Clusters Associated with Luminous IRAS Point Sources}",
      journal = {\apj},
         year = 1993,
        month = apr,
       volume = {407},
        pages = {657},
          doi = {10.1086/172548},
       adsurl = {https://ui.adsabs.harvard.edu/abs/1993ApJ...407..657C}
}

@article{10.1093/mnras/stv2248,
    author = {Marsh, K. A. and Whitworth, A. P. and Lomax, O.},
    title = {Temperature as a third dimension in column-density mapping of dusty astrophysical structures associated with star formation},
    journal = {Monthly Notices of the Royal Astronomical Society},
    volume = {454},
    number = {4},
    pages = {4282-4292},
    year = {2015},
    month = {10},
    issn = {0035-8711},
    doi = {10.1093/mnras/stv2248},
    url = {https://doi.org/10.1093/mnras/stv2248},
    eprint = {https://academic.oup.com/mnras/article-pdf/454/4/4282/18510384/stv2248.pdf},
}

@ARTICLE{2008AJ....135..637H,
       author = {{Herbig}, G.~H.},
        title = "{History and Spectroscopy of EXor Candidates}",
      journal = {\aj},
         year = 2008,
        month = feb,
       volume = {135},
       number = {2},
        pages = {637-648},
          doi = {10.1088/0004-6256/135/2/637},
       adsurl = {https://ui.adsabs.harvard.edu/abs/2008AJ....135..637H}
}

@ARTICLE{1988MNRAS.235..441H,
       author = {{Harris}, Stella and {Clegg}, Peter and {Hughes}, Joanne},
        title = "{T Tauri stars in Taurus -the IRAS view.}",
      journal = {\mnras},
         year = 1988,
        month = nov,
       volume = {235},
        pages = {441-456},
          doi = {10.1093/mnras/235.2.441},
       adsurl = {https://ui.adsabs.harvard.edu/abs/1988MNRAS.235..441H}
}

@ARTICLE{1977ApJ...217..693H,
       author = {{Herbig}, G.~H.},
        title = "{Eruptive phenomena in early stellar evolution.}",
      journal = {\apj},
         year = 1977,
        month = nov,
       volume = {217},
        pages = {693-715},
          doi = {10.1086/155615},
       adsurl = {https://ui.adsabs.harvard.edu/abs/1977ApJ...217..693H}
}

@ARTICLE{1996ARA&A..34..207H,
       author = {{Hartmann}, Lee and {Kenyon}, Scott J.},
        title = "{The FU Orionis Phenomenon}",
      journal = {\araa},
         year = 1996,
        month = jan,
       volume = {34},
        pages = {207-240},
          doi = {10.1146/annurev.astro.34.1.207},
       adsurl = {https://ui.adsabs.harvard.edu/abs/1996ARA&A..34..207H}
}

@INPROCEEDINGS{2014prpl.conf..387A,
       author = {{Audard}, M. and {{\'A}brah{\'a}m}, P. and {Dunham}, M.~M. and {Green}, J.~D. and {Grosso}, N. and {Hamaguchi}, K. and {Kastner}, J.~H. and {K{\'o}sp{\'a}l}, {\'A}. and {Lodato}, G. and {Romanova}, M.~M. and {Skinner}, S.~L. and {Vorobyov}, E.~I. and {Zhu}, Z.},
        title = "{Episodic Accretion in Young Stars}",
    booktitle = {Protostars and Planets VI},
         year = 2014,
       editor = {{Beuther}, Henrik and {Klessen}, Ralf S. and {Dullemond}, Cornelis P. and {Henning}, Thomas},
        month = jan,
        pages = {387-410},
          doi = {10.2458/azu_uapress_9780816531240-ch017},
archivePrefix = {arXiv},
       eprint = {1401.3368},
 primaryClass = {astro-ph.SR},
       adsurl = {https://ui.adsabs.harvard.edu/abs/2014prpl.conf..387A}
}

@ARTICLE{2023A&A...674A..41G,
       author = {{Gaia Collaboration} and {Bailer-Jones}, C.~A.~L. and {Teyssier}, D. and {Delchambre}, L. and {Ducourant}, C. and {Garabato}, D. and {Hatzidimitriou}, D. and {Klioner}, S.~A. and {Rimoldini}, L. and {Bellas-Velidis}, I. and {Carballo}, R. and {Carnerero}, M.~I. and {Diener}, C. and {Fouesneau}, M. and {Galluccio}, L. and {Gavras}, P. and {Krone-Martins}, A. and {Raiteri}, C.~M. and {Teixeira}, R. and {Brown}, A.~G.~A. and {Vallenari}, A. and {Prusti}, T. and {de Bruijne}, J.~H.~J. and {Arenou}, F. and {Babusiaux}, C. and {Biermann}, M. and {Creevey}, O.~L. and {Evans}, D.~W. and {Eyer}, L. and {Guerra}, R. and {Hutton}, A. and {Jordi}, C. and {Lammers}, U.~L. and {Lindegren}, L. and {Luri}, X. and {Mignard}, F. and {Panem}, C. and {Pourbaix}, D. and {Randich}, S. and {Sartoretti}, P. and {Soubiran}, C. and {Tanga}, P. and {Walton}, N.~A. and {Bastian}, U. and {Drimmel}, R. and {Jansen}, F. and {Katz}, D. and {Lattanzi}, M.~G. and {van Leeuwen}, F. and {Bakker}, J. and {Cacciari}, C. and {Casta{\~n}eda}, J. and {De Angeli}, F. and {Fabricius}, C. and {Fr{\'e}mat}, Y. and {Guerrier}, A. and {Heiter}, U. and {Masana}, E. and {Messineo}, R. and {Mowlavi}, N. and {Nicolas}, C. and {Nienartowicz}, K. and {Pailler}, F. and {Panuzzo}, P. and {Riclet}, F. and {Roux}, W. and {Seabroke}, G.~M. and {Sordo}, R. and {Th{\'e}venin}, F. and {Gracia-Abril}, G. and {Portell}, J. and {Altmann}, M. and {Andrae}, R. and {Audard}, M. and {Benson}, K. and {Berthier}, J. and {Blomme}, R. and {Burgess}, P.~W. and {Busonero}, D. and {Busso}, G. and {C{\'a}novas}, H. and {Carry}, B. and {Cellino}, A. and {Cheek}, N. and {Clementini}, G. and {Damerdji}, Y. and {Davidson}, M. and {de Teodoro}, P. and {Nu{\~n}ez Campos}, M. and {Dell'Oro}, A. and {Esquej}, P. and {Fern{\'a}ndez-Hern{\'a}ndez}, J. and {Fraile}, E. and {Garc{\'\i}a-Lario}, P. and {Gosset}, E. and {Haigron}, R. and {Halbwachs}, J.-L. and {Hambly}, N.~C. and {Harrison}, D.~L. and {Hern{\'a}ndez}, J. and {Hestroffer}, D. and {Hodgkin}, S.~T. and {Holl}, B. and {Jan{\ss}en}, K. and {Jevardat de Fombelle}, G. and {Jordan}, S. and {Lanzafame}, A.~C. and {L{\"o}ffler}, W. and {Marchal}, O. and {Marrese}, P.~M. and {Moitinho}, A. and {Muinonen}, K. and {Osborne}, P. and {Pancino}, E. and {Pauwels}, T. and {Recio-Blanco}, A. and {Reyl{\'e}}, C. and {Riello}, M. and {Roegiers}, T. and {Rybizki}, J. and {Sarro}, L.~M. and {Siopis}, C. and {Smith}, M. and {Sozzetti}, A. and {Utrilla}, E. and {van Leeuwen}, M. and {Abbas}, U. and {{\'A}brah{\'a}m}, P. and {Abreu Aramburu}, A. and {Aerts}, C. and {Aguado}, J.~J. and {Ajaj}, M. and {Aldea-Montero}, F. and {Altavilla}, G. and {{\'A}lvarez}, M.~A. and {Alves}, J. and {Anderson}, R.~I. and {Anglada Varela}, E. and {Antoja}, T. and {Baines}, D. and {Baker}, S.~G. and {Balaguer-N{\'u}{\~n}ez}, L. and {Balbinot}, E. and {Balog}, Z. and {Barache}, C. and {Barbato}, D. and {Barros}, M. and {Barstow}, M.~A. and {Bartolom{\'e}}, S. and {Bassilana}, J.-L. and {Bauchet}, N. and {Becciani}, U. and {Bellazzini}, M. and {Berihuete}, A. and {Bernet}, M. and {Bertone}, S. and {Bianchi}, L. and {Binnenfeld}, A. and {Blanco-Cuaresma}, S. and {Boch}, T. and {Bombrun}, A. and {Bossini}, D. and {Bouquillon}, S. and {Bragaglia}, A. and {Bramante}, L. and {Breedt}, E. and {Bressan}, A. and {Brouillet}, N. and {Brugaletta}, E. and {Bucciarelli}, B. and {Burlacu}, A. and {Butkevich}, A.~G. and {Buzzi}, R. and {Caffau}, E. and {Cancelliere}, R. and {Cantat-Gaudin}, T. and {Carlucci}, T. and {Carrasco}, J.~M. and {Casamiquela}, L. and {Castellani}, M. and {Castro-Ginard}, A. and {Chaoul}, L. and {Charlot}, P. and {Chemin}, L. and {Chiaramida}, V. and {Chiavassa}, A. and {Chornay}, N. and {Comoretto}, G. and {Contursi}, G. and {Cooper}, W.~J. and {Cornez}, T. and {Cowell}, S. and {Crifo}, F. and {Cropper}, M. and {Crosta}, M. and {Crowley}, C. and {Dafonte}, C. and {Dapergolas}, A. and {David}, P. and {de Laverny}, P.},
        title = "{Gaia Data Release 3. The extragalactic content}",
      journal = {\aap},
         year = 2023,
        month = jun,
       volume = {674},
          eid = {A41},
        pages = {A41},
          doi = {10.1051/0004-6361/202243232},
archivePrefix = {arXiv},
       eprint = {2206.05681},
 primaryClass = {astro-ph.GA},
       adsurl = {https://ui.adsabs.harvard.edu/abs/2023A&A...674A..41G}
}

@ARTICLE{2017A&A...600A..11R,
       author = {{Robitaille}, T.~P.},
        title = "{A modular set of synthetic spectral energy distributions for young stellar objects}",
      journal = {\aap},
         year = 2017,
        month = apr,
       volume = {600},
          eid = {A11},
        pages = {A11},
          doi = {10.1051/0004-6361/201425486},
archivePrefix = {arXiv},
       eprint = {1703.05765},
 primaryClass = {astro-ph.SR},
       adsurl = {https://ui.adsabs.harvard.edu/abs/2017A&A...600A..11R}
}

@dataset{2019ipac.data...I1W,
       author = {{Wright}, Edward L. and {Eisenhardt}, Peter R.~M. and {Mainzer}, Amy K. and {Ressler}, Michael E. and {Cutri}, Roc M. and {Jarrett}, Thomas and {Kirkpatrick}, J. Davy and {Padgett}, Deborah and {McMillan}, Robert S. and {Skrutskie}, Michael and {Stanford}, S.~A. and {Cohen}, Martin and {Walker}, Russell G. and {Mather}, John C. and {Leisawitz}, David and {Gautier}, III, Thomas N. and {McLean}, Ian and {Benford}, Dominic and {Lonsdale}, Carol J. and {Blain}, Andrew and {Mendez}, Bryan and {Irace}, William R. and {Duval}, Valerie and {Liu}, Fengchuan and {Royer}, Don and {Heinrichsen}, Ingolf and {Howard}, Joan and {Shannon}, Mark and {Kendall}, Martha and {Walsh}, Amy L. and {Larsen}, Mark and {Cardon}, Joel G. and {Schick}, Scott and {Schwalm}, Mark and {Abid}, Mohamed and {Fabinsky}, Beth and {Naes}, Larry and {Tsai}, ChaoWei},
        title = "{AllWISE Source Catalog}",
 howpublished = {NASA IPAC DataSet, IRSA1},
         year = 2019,
        month = jan,
          doi = {10.26131/IRSA1},
       adsurl = {https://ui.adsabs.harvard.edu/abs/2019ipac.data...I1W}
}

@ARTICLE{2006MNRAS.367..454H,
       author = {{Hewett}, P.~C. and {Warren}, S.~J. and {Leggett}, S.~K. and {Hodgkin}, S.~T.},
        title = "{The UKIRT Infrared Deep Sky Survey ZY JHK photometric system: passbands and synthetic colours}",
      journal = {\mnras},
         year = 2006,
        month = apr,
       volume = {367},
       number = {2},
        pages = {454-468},
          doi = {10.1111/j.1365-2966.2005.09969.x},
archivePrefix = {arXiv},
       eprint = {astro-ph/0601592},
 primaryClass = {astro-ph},
       adsurl = {https://ui.adsabs.harvard.edu/abs/2006MNRAS.367..454H}
}

@article{1993AJ....105.1927G,
       author = {{Gomez}, M. and {Hartmann}, L. and {Kenyon}, S.~J. and {Hewett}, R.},
        title = "{On the Spatial Distribution of pre-Main-Sequence Stars in Taurus}",
      journal = {\aj},
         year = 1993,
        month = may,
       volume = {105},
        pages = {1927},
          doi = {10.1086/116567},
       adsurl = {https://ui.adsabs.harvard.edu/abs/1993AJ....105.1927G}
}

@article{2006AJ....131.1163S,
       author = {{Skrutskie}, M.~F. and {Cutri}, R.~M. and {Stiening}, R. and {Weinberg}, M.~D. and {Schneider}, S. and {Carpenter}, J.~M. and {Beichman}, C. and {Capps}, R. and {Chester}, T. and {Elias}, J. and {Huchra}, J. and {Liebert}, J. and {Lonsdale}, C. and {Monet}, D.~G. and {Price}, S. and {Seitzer}, P. and {Jarrett}, T. and {Kirkpatrick}, J.~D. and {Gizis}, J.~E. and {Howard}, E. and {Evans}, T. and {Fowler}, J. and {Fullmer}, L. and {Hurt}, R. and {Light}, R. and {Kopan}, E.~L. and {Marsh}, K.~A. and {McCallon}, H.~L. and {Tam}, R. and {Van Dyk}, S. and {Wheelock}, S.},
        title = "{The Two Micron All Sky Survey (2MASS)}",
      journal = {\aj},
         year = 2006,
        month = feb,
       volume = {131},
       number = {2},
        pages = {1163-1183},
          doi = {10.1086/498708},
       adsurl = {https://ui.adsabs.harvard.edu/abs/2006AJ....131.1163S}
}

@article{2008MNRAS.391..136L,
       author = {{Lucas}, P.~W. and {Hoare}, M.~G. and {Longmore}, A. and {Schr{\"o}der}, A.~C. and {Davis}, C.~J. and {Adamson}, A. and {Bandyopadhyay}, R.~M. and {de Grijs}, R. and {Smith}, M. and {Gosling}, A. and {Mitchison}, S. and {G{\'a}sp{\'a}r}, A. and {Coe}, M. and {Tamura}, M. and {Parker}, Q. and {Irwin}, M. and {Hambly}, N. and {Bryant}, J. and {Collins}, R.~S. and {Cross}, N. and {Evans}, D.~W. and {Gonzalez-Solares}, E. and {Hodgkin}, S. and {Lewis}, J. and {Read}, M. and {Riello}, M. and {Sutorius}, E.~T.~W. and {Lawrence}, A. and {Drew}, J.~E. and {Dye}, S. and {Thompson}, M.~A.},
        title = "{The UKIDSS Galactic Plane Survey}",
      journal = {\mnras},
         year = 2008,
        month = nov,
       volume = {391},
       number = {1},
        pages = {136-163},
          doi = {10.1111/j.1365-2966.2008.13924.x},
archivePrefix = {arXiv},
       eprint = {0712.0100},
 primaryClass = {astro-ph},
       adsurl = {https://ui.adsabs.harvard.edu/abs/2008MNRAS.391..136L}
}

@article{2015MNRAS.447.2307M,
       author = {{Mallick}, K.~K. and {Ojha}, D.~K. and {Tamura}, M. and {Linz}, H. and {Samal}, M.~R. and {Ghosh}, S.~K.},
        title = "{Study of morphology and stellar content of the Galactic H II region IRAS 16148-5011}",
      journal = {\mnras},
         year = 2015,
        month = mar,
       volume = {447},
       number = {3},
        pages = {2307-2321},
          doi = {10.1093/mnras/stu2584},
archivePrefix = {arXiv},
       eprint = {1412.1651},
 primaryClass = {astro-ph.GA},
       adsurl = {https://ui.adsabs.harvard.edu/abs/2015MNRAS.447.2307M}
}

@article{10.1111/j.1365-2966.2012.21948.x,
    author = {Bressan, Alessandro and Marigo, Paola and Girardi, Léo and Salasnich, Bernardo and Dal Cero, Claudia and Rubele, Stefano and Nanni, Ambra},
    title = {PARSEC: stellar tracks and isochrones with the PAdova and TRieste Stellar Evolution Code},
    journal = {Monthly Notices of the Royal Astronomical Society},
    volume = {427},
    number = {1},
    pages = {127-145},
    year = {2012},
    month = {11},
    issn = {0035-8711},
    doi = {10.1111/j.1365-2966.2012.21948.x},
    url = {https://doi.org/10.1111/j.1365-2966.2012.21948.x},
    eprint = {https://academic.oup.com/mnras/article-pdf/427/1/127/18231042/427-1-127.pdf},
}

@article{2015A&A...577A..42B,
       author = {{Baraffe}, Isabelle and {Homeier}, Derek and {Allard}, France and {Chabrier}, Gilles},
        title = "{New evolutionary models for pre-main sequence and main sequence low-mass stars down to the hydrogen-burning limit}",
      journal = {\aap},
         year = 2015,
        month = may,
       volume = {577},
          eid = {A42},
        pages = {A42},
          doi = {10.1051/0004-6361/201425481},
archivePrefix = {arXiv},
       eprint = {1503.04107},
 primaryClass = {astro-ph.SR},
       adsurl = {https://ui.adsabs.harvard.edu/abs/2015A&A...577A..42B}
}

@article{2011PASP..123.1218A,
       author = {{Aniano}, G. and {Draine}, B.~T. and {Gordon}, K.~D. and {Sandstrom}, K.},
        title = "{Common-Resolution Convolution Kernels for Space- and Ground-Based Telescopes}",
      journal = {\pasp},
         year = 2011,
        month = oct,
       volume = {123},
       number = {908},
        pages = {1218},
          doi = {10.1086/662219},
archivePrefix = {arXiv},
       eprint = {1106.5065},
 primaryClass = {astro-ph.IM},
       adsurl = {https://ui.adsabs.harvard.edu/abs/2011PASP..123.1218A}
}

@article{2013A&A...551A..98L,
       author = {{Launhardt}, R. and {Stutz}, A.~M. and {Schmiedeke}, A. and {Henning}, Th. and {Krause}, O. and {Balog}, Z. and {Beuther}, H. and {Birkmann}, S. and {Hennemann}, M. and {Kainulainen}, J. and {Khanzadyan}, T. and {Linz}, H. and {Lippok}, N. and {Nielbock}, M. and {Pitann}, J. and {Ragan}, S. and {Risacher}, C. and {Schmalzl}, M. and {Shirley}, Y.~L. and {Stecklum}, B. and {Steinacker}, J. and {Tackenberg}, J.},
        title = "{The Earliest Phases of Star Formation (EPoS): a Herschel key project. The thermal structure of low-mass molecular cloud cores}",
      journal = {\aap},
         year = 2013,
        month = mar,
       volume = {551},
          eid = {A98},
        pages = {A98},
          doi = {10.1051/0004-6361/201220477},
archivePrefix = {arXiv},
       eprint = {1301.1498},
 primaryClass = {astro-ph.SR},
       adsurl = {https://ui.adsabs.harvard.edu/abs/2013A&A...551A..98L}
}

@article{2001A&A...377.1023L,
       author = {{Lombardi}, M. and {Alves}, J.},
        title = "{Mapping the interstellar dust with near-infrared observations: An optimized multi-band technique}",
      journal = {\aap},
         year = 2001,
        month = oct,
       volume = {377},
        pages = {1023-1034},
          doi = {10.1051/0004-6361:20011099},
archivePrefix = {arXiv},
       eprint = {astro-ph/0109135},
 primaryClass = {astro-ph},
       adsurl = {https://ui.adsabs.harvard.edu/abs/2001A&A...377.1023L}
}

@article{Li_2024,
doi = {10.3847/1538-4357/ad2a59},
url = {https://dx.doi.org/10.3847/1538-4357/ad2a59},
year = {2024},
month = {apr},
publisher = {The American Astronomical Society},
volume = {965},
number = {1},
pages = {29},
author = {Li, Jun and Jiang, Biwei and Zhao, He and Chen, Xi and Yang, Yang},
title = {Spatial Variations of Dust Opacity and Grain Growth in Dark Clouds: L1689, L1709, and L1712},
journal = {The Astrophysical Journal}
}

@article{Chu_2021,
   title={Constraining Spatial Densities of Early Ice Formation in Small Dense Molecular Cores from Extinction Maps},
   volume={918},
   ISSN={1538-4357},
   url={http://dx.doi.org/10.3847/1538-4357/ac0ae8},
   DOI={10.3847/1538-4357/ac0ae8},
   number={1},
   journal={The Astrophysical Journal},
   publisher={American Astronomical Society},
   author={Chu, Laurie E. U. and Hodapp, Klaus W.},
   year={2021},
   month=aug, pages={2} }

@article{1997AJ....114..288M,
       author = {{Meyer}, Michael R. and {Calvet}, Nuria and {Hillenbrand}, Lynne A.},
        title = "{Intrinsic Near-Infrared Excesses of T Tauri Stars: Understanding the Classical T Tauri Star Locus}",
      journal = {\aj},
         year = 1997,
        month = jul,
       volume = {114},
        pages = {288-300},
          doi = {10.1086/118474},
       adsurl = {https://ui.adsabs.harvard.edu/abs/1997AJ....114..288M}
}

@article{1988PASP..100.1134B,
       author = {{Bessell}, M.~S. and {Brett}, J.~M.},
        title = "{JHKLM Photometry: Standard Systems, Passbands, and Intrinsic Colors}",
      journal = {\pasp},
         year = 1988,
        month = sep,
       volume = {100},
        pages = {1134},
          doi = {10.1086/132281},
       adsurl = {https://ui.adsabs.harvard.edu/abs/1988PASP..100.1134B}
}

@article{2009ApJS..184...18G,
       author = {{Gutermuth}, R.~A. and {Megeath}, S.~T. and {Myers}, P.~C. and {Allen}, L.~E. and {Pipher}, J.~L. and {Fazio}, G.~G.},
        title = "{A Spitzer Survey of Young Stellar Clusters Within One Kiloparsec of the Sun: Cluster Core Extraction and Basic Structural Analysis}",
      journal = {\apjs},
         year = 2009,
        month = sep,
       volume = {184},
       number = {1},
        pages = {18-83},
          doi = {10.1088/0067-0049/184/1/18},
archivePrefix = {arXiv},
       eprint = {0906.0201},
 primaryClass = {astro-ph.SR},
       adsurl = {https://ui.adsabs.harvard.edu/abs/2009ApJS..184...18G}
}

@article{2014ApJ...791..131K,
       author = {{Koenig}, X.~P. and {Leisawitz}, D.~T.},
        title = "{A Classification Scheme for Young Stellar Objects Using the Wide-field Infrared Survey Explorer AllWISE Catalog: Revealing Low-density Star Formation in the Outer Galaxy}",
      journal = {\apj},
         year = 2014,
        month = aug,
       volume = {791},
       number = {2},
          eid = {131},
        pages = {131},
          doi = {10.1088/0004-637X/791/2/131},
archivePrefix = {arXiv},
       eprint = {1407.2262},
 primaryClass = {astro-ph.GA},
       adsurl = {https://ui.adsabs.harvard.edu/abs/2014ApJ...791..131K}
}

@article{Bisht_2020,
   title={An Investigation of Poorly Studied Open Cluster NGC 4337 Using Multicolor Photometric and Gaia DR2 Astrometric Data},
   volume={160},
   ISSN={1538-3881},
   url={http://dx.doi.org/10.3847/1538-3881/ab9ffd},
   DOI={10.3847/1538-3881/ab9ffd},
   number={3},
   journal={The Astronomical Journal},
   publisher={American Astronomical Society},
   author={Bisht, D. and Elsanhoury, W. H. and Zhu, Qingfeng and Sariya, Devesh P. and Yadav, R. K. S. and Rangwal, Geeta and Durgapal, Alok and Jiang, Ing-Guey},
   year={2020},
   month=aug, pages={119} }

@ARTICLE{2006ApJS..167..256R,
       author = {{Robitaille}, Thomas P. and {Whitney}, Barbara A. and {Indebetouw}, Remy and {Wood}, Kenneth and {Denzmore}, Pia},
        title = "{Interpreting Spectral Energy Distributions from Young Stellar Objects. I. A Grid of 200,000 YSO Model SEDs}",
      journal = {\apjs},
         year = 2006,
        month = dec,
       volume = {167},
       number = {2},
        pages = {256-285},
          doi = {10.1086/508424},
archivePrefix = {arXiv},
       eprint = {astro-ph/0608234},
 primaryClass = {astro-ph},
       adsurl = {https://ui.adsabs.harvard.edu/abs/2006ApJS..167..256R}
}

@ARTICLE{2003ApJ...591.1049W,
       author = {{Whitney}, Barbara A. and {Wood}, Kenneth and {Bjorkman}, J.~E. and {Wolff}, Michael J.},
        title = "{Two-dimensional Radiative Transfer in Protostellar Envelopes. I. Effects of Geometry on Class I Sources}",
      journal = {\apj},
         year = 2003,
        month = jul,
       volume = {591},
       number = {2},
        pages = {1049-1063},
          doi = {10.1086/375415},
archivePrefix = {arXiv},
       eprint = {astro-ph/0303479},
 primaryClass = {astro-ph},
       adsurl = {https://ui.adsabs.harvard.edu/abs/2003ApJ...591.1049W}
}

@ARTICLE{1981ApJ...249..481C,
       author = {{Cohen}, J.~G. and {Frogel}, J.~A. and {Persson}, S.~E. and {Elias}, J.~H.},
        title = "{Bolometric luminosities and infrared properties of carbon stars in the Magellanic Clouds and the galaxy.}",
      journal = {\apj},
         year = 1981,
        month = oct,
       volume = {249},
        pages = {481-503},
          doi = {10.1086/159308},
       adsurl = {https://ui.adsabs.harvard.edu/abs/1981ApJ...249..481C}
}

@ARTICLE{2003ApJ...598.1079W,
       author = {{Whitney}, Barbara A. and {Wood}, Kenneth and {Bjorkman}, J.~E. and {Cohen}, Martin},
        title = "{Two-dimensional Radiative Transfer in Protostellar Envelopes. II. An Evolutionary Sequence}",
      journal = {\apj},
         year = 2003,
        month = dec,
       volume = {598},
       number = {2},
        pages = {1079-1099},
          doi = {10.1086/379068},
archivePrefix = {arXiv},
       eprint = {astro-ph/0309007},
 primaryClass = {astro-ph},
       adsurl = {https://ui.adsabs.harvard.edu/abs/2003ApJ...598.1079W}
}

@article{2007ApJS..169..328R,
       author = {{Robitaille}, Thomas P. and {Whitney}, Barbara A. and {Indebetouw}, Remy and {Wood}, Kenneth},
        title = "{Interpreting Spectral Energy Distributions from Young Stellar Objects. II. Fitting Observed SEDs Using a Large Grid of Precomputed Models}",
      journal = {\apjs},
         year = 2007,
        month = apr,
       volume = {169},
       number = {2},
        pages = {328-352},
          doi = {10.1086/512039},
archivePrefix = {arXiv},
       eprint = {astro-ph/0612690},
 primaryClass = {astro-ph},
       adsurl = {https://ui.adsabs.harvard.edu/abs/2007ApJS..169..328R}
}

@article{2020MNRAS.494.5851S,
       author = {{Saha}, Piyali and {Gopinathan}, Maheswar and {Kamath}, Umanath and {Lee}, Chang Won and {Manoj}, P. and {Mathew}, Blesson and {Sharma}, Ekta},
        title = "{A census of young stellar population associated with the Herbig Be star HD 200775}",
      journal = {\mnras},
         year = 2020,
        month = jun,
       volume = {494},
       number = {4},
        pages = {5851-5871},
          doi = {10.1093/mnras/staa1053},
archivePrefix = {arXiv},
       eprint = {2005.00519},
 primaryClass = {astro-ph.SR},
       adsurl = {https://ui.adsabs.harvard.edu/abs/2020MNRAS.494.5851S}
}

@ARTICLE{2007ApJ...663.1069F,
       author = {{Flaherty}, K.~M. and {Pipher}, J.~L. and {Megeath}, S.~T. and {Winston}, E.~M. and {Gutermuth}, R.~A. and {Muzerolle}, J. and {Allen}, L.~E. and {Fazio}, G.~G.},
        title = "{Infrared Extinction toward Nearby Star-forming Regions}",
      journal = {\apj},
         year = 2007,
        month = jul,
       volume = {663},
       number = {2},
        pages = {1069-1082},
          doi = {10.1086/518411},
archivePrefix = {arXiv},
       eprint = {astro-ph/0703777},
 primaryClass = {astro-ph},
       adsurl = {https://ui.adsabs.harvard.edu/abs/2007ApJ...663.1069F}
}

@ARTICLE{2004MNRAS.348..589C,
       author = {{Cartwright}, Annabel and {Whitworth}, Anthony P.},
        title = "{The statistical analysis of star clusters}",
      journal = {\mnras},
         year = 2004,
        month = feb,
       volume = {348},
       number = {2},
        pages = {589-598},
          doi = {10.1111/j.1365-2966.2004.07360.x},
archivePrefix = {arXiv},
       eprint = {astro-ph/0403474},
 primaryClass = {astro-ph},
       adsurl = {https://ui.adsabs.harvard.edu/abs/2004MNRAS.348..589C}
}

@ARTICLE{2006A&A...449..151S,
       author = {{Schmeja}, S. and {Klessen}, R.~S.},
        title = "{Evolving structures of star-forming clusters}",
      journal = {\aap},
         year = 2006,
        month = apr,
       volume = {449},
       number = {1},
        pages = {151-159},
          doi = {10.1051/0004-6361:20054464},
archivePrefix = {arXiv},
       eprint = {astro-ph/0511448},
 primaryClass = {astro-ph},
       adsurl = {https://ui.adsabs.harvard.edu/abs/2006A&A...449..151S}
}

@ARTICLE{1988ApJ...328..143L,
       author = {{Lada}, Charles J. and {Margulis}, Michael and {Sofue}, Yoshiaki and {Nakai}, Naomasa and {Handa}, Toshihiro},
        title = "{Observations of Molecular and Atomic Clouds in M31}",
      journal = {\apj},
         year = 1988,
        month = may,
       volume = {328},
        pages = {143},
          doi = {10.1086/166275},
       adsurl = {https://ui.adsabs.harvard.edu/abs/1988ApJ...328..143L}
}

@article{Gutermuth_2005,
   title={The Initial Configuration of Young Stellar Clusters: AK‐Band Number Counts Analysis of the Surface Density of Stars},
   volume={632},
   ISSN={1538-4357},
   url={http://dx.doi.org/10.1086/432460},
   DOI={10.1086/432460},
   number={1},
   journal={The Astrophysical Journal},
   publisher={American Astronomical Society},
   author={Gutermuth, Robert A. and Megeath, S. Thomas and Pipher, Judith L. and Williams, Jonathan P. and Allen, Lori E. and Myers, Philip C. and Raines, S. Nicholas},
   year={2005},
   month=oct, pages={397–420} }

@article{10.1093/mnras/staa2412,
    author = {Sharma, Saurabh and Ghosh, Arpan and Ojha, D K and Pandey, R and Sinha, T and Pandey, A K and Ghosh, S K and Panwar, N and Pandey, S B},
    title = {The disintegrating old open cluster Czernik 3},
    journal = {Monthly Notices of the Royal Astronomical Society},
    volume = {498},
    number = {2},
    pages = {2309-2322},
    year = {2020},
    month = {09},
    issn = {0035-8711},
    doi = {10.1093/mnras/staa2412},
    url = {https://doi.org/10.1093/mnras/staa2412},
    eprint = {https://academic.oup.com/mnras/article-pdf/498/2/2309/33777039/staa2412.pdf},
}

@ARTICLE{1985ApJ...298...80C,
       author = {{Casertano}, S. and {Hut}, P.},
        title = "{Core radius and density measurements in N-body experiments Connections with theoretical and observational definitions}",
      journal = {\apj},
         year = 1985,
        month = nov,
       volume = {298},
        pages = {80-94},
          doi = {10.1086/163589},
       adsurl = {https://ui.adsabs.harvard.edu/abs/1985ApJ...298...80C}
}

@article{10.1111/j.1365-2966.2007.12064.x,
    author = {Bastian, N. and Ercolano, B. and Gieles, M. and Rosolowsky, E. and Scheepmaker, R. A. and Gutermuth, R. and Efremov, Yu.},
    title = {Hierarchical star formation in M33: fundamental properties of the star-forming regions},
    journal = {Monthly Notices of the Royal Astronomical Society},
    volume = {379},
    number = {4},
    pages = {1302-1312},
    year = {2007},
    month = {08},
    issn = {0035-8711},
    doi = {10.1111/j.1365-2966.2007.12064.x},
    url = {https://doi.org/10.1111/j.1365-2966.2007.12064.x},
    eprint = {https://academic.oup.com/mnras/article-pdf/379/4/1302/3174457/mnras0379-1302.pdf},
}

@ARTICLE{1994ApJ...429..694L,
       author = {{Lada}, Charles J. and {Lada}, Elizabeth A. and {Clemens}, Dan P. and {Bally}, John},
        title = "{Dust Extinction and Molecular Gas in the Dark Cloud IC 5146}",
      journal = {\apj},
         year = 1994,
        month = jul,
       volume = {429},
        pages = {694},
          doi = {10.1086/174354},
       adsurl = {https://ui.adsabs.harvard.edu/abs/1994ApJ...429..694L}
}

@ARTICLE{2009ApJ...703...52L,
       author = {{Lada}, Charles J. and {Lombardi}, Marco and {Alves}, Jo{\~a}o F.},
        title = "{The California Molecular Cloud}",
      journal = {\apj},
         year = 2009,
        month = sep,
       volume = {703},
       number = {1},
        pages = {52-59},
          doi = {10.1088/0004-637X/703/1/52},
archivePrefix = {arXiv},
       eprint = {0908.0646},
 primaryClass = {astro-ph.GA},
       adsurl = {https://ui.adsabs.harvard.edu/abs/2009ApJ...703...52L}
}

@ARTICLE{2006A&A...454..781L,
       author = {{Lombardi}, M. and {Alves}, J. and {Lada}, C.~J.},
        title = "{2MASS wide field extinction maps. I. The Pipe nebula}",
      journal = {\aap},
         year = 2006,
        month = aug,
       volume = {454},
       number = {3},
        pages = {781-796},
          doi = {10.1051/0004-6361:20042474},
archivePrefix = {arXiv},
       eprint = {astro-ph/0606670},
 primaryClass = {astro-ph},
       adsurl = {https://ui.adsabs.harvard.edu/abs/2006A&A...454..781L}
}

@ARTICLE{2005A&A...436..895G,
       author = {{Girardi}, L. and {Groenewegen}, M.~A.~T. and {Hatziminaoglou}, E. and {da Costa}, L.},
        title = "{Star counts in the Galaxy. Simulating from very deep to very shallow photometric surveys with the TRILEGAL code}",
      journal = {\aap},
         year = 2005,
        month = jun,
       volume = {436},
       number = {3},
        pages = {895-915},
          doi = {10.1051/0004-6361:20042352},
archivePrefix = {arXiv},
       eprint = {astro-ph/0504047},
 primaryClass = {astro-ph},
       adsurl = {https://ui.adsabs.harvard.edu/abs/2005A&A...436..895G}
}

@ARTICLE{1998ApJ...500..525S,
       author = {{Schlegel}, David J. and {Finkbeiner}, Douglas P. and {Davis}, Marc},
        title = "{Maps of Dust Infrared Emission for Use in Estimation of Reddening and Cosmic Microwave Background Radiation Foregrounds}",
      journal = {\apj},
         year = 1998,
        month = jun,
       volume = {500},
       number = {2},
        pages = {525-553},
          doi = {10.1086/305772},
archivePrefix = {arXiv},
       eprint = {astro-ph/9710327},
 primaryClass = {astro-ph},
       adsurl = {https://ui.adsabs.harvard.edu/abs/1998ApJ...500..525S}
}

@ARTICLE{2011A&A...535A.128B,
       author = {{Battersby}, C. and {Bally}, J. and {Ginsburg}, A. and {Bernard}, J. -P. and {Brunt}, C. and {Fuller}, G.~A. and {Martin}, P. and {Molinari}, S. and {Mottram}, J. and {Peretto}, N. and {Testi}, L. and {Thompson}, M.~A.},
        title = "{Characterizing precursors to stellar clusters with Herschel}",
      journal = {\aap},
         year = 2011,
        month = nov,
       volume = {535},
          eid = {A128},
        pages = {A128},
          doi = {10.1051/0004-6361/201116559},
archivePrefix = {arXiv},
       eprint = {1101.4654},
 primaryClass = {astro-ph.GA},
       adsurl = {https://ui.adsabs.harvard.edu/abs/2011A&A...535A.128B}
}

@ARTICLE{2014A&A...566A..45L,
       author = {{Lombardi}, Marco and {Bouy}, Herv{\'e} and {Alves}, Jo{\~a}o and {Lada}, Charles J.},
        title = "{Herschel-Planck dust optical-depth and column-density maps. I. Method description and results for Orion}",
      journal = {\aap},
         year = 2014,
        month = jun,
       volume = {566},
          eid = {A45},
        pages = {A45},
          doi = {10.1051/0004-6361/201323293},
archivePrefix = {arXiv},
       eprint = {1404.0032},
 primaryClass = {astro-ph.SR},
       adsurl = {https://ui.adsabs.harvard.edu/abs/2014A&A...566A..45L}
}

@ARTICLE{2012A&A...540A..10S,
       author = {{Sadavoy}, S.~I. and {di Francesco}, J. and {Andr{\'e}}, Ph. and {Pezzuto}, S. and {Bernard}, J. -P. and {Bontemps}, S. and {Bressert}, E. and {Chitsazzadeh}, S. and {Fallscheer}, C. and {Hennemann}, M. and {Hill}, T. and {Martin}, P. and {Motte}, F. and {Nguyen Luong}, Q. and {Peretto}, N. and {Reid}, M. and {Schneider}, N. and {Testi}, L. and {White}, G.~J. and {Wilson}, C.},
        title = "{Herschel observations of a potential core-forming clump: Perseus B1-E}",
      journal = {\aap},
         year = 2012,
        month = apr,
       volume = {540},
          eid = {A10},
        pages = {A10},
          doi = {10.1051/0004-6361/201117934},
archivePrefix = {arXiv},
       eprint = {1111.7021},
 primaryClass = {astro-ph.GA},
       adsurl = {https://ui.adsabs.harvard.edu/abs/2012A&A...540A..10S}
}

@ARTICLE{2012A&A...547A..11N,
       author = {{Nielbock}, M. and {Launhardt}, R. and {Steinacker}, J. and {Stutz}, A.~M. and {Balog}, Z. and {Beuther}, H. and {Bouwman}, J. and {Henning}, Th. and {Hily-Blant}, P. and {Kainulainen}, J. and {Krause}, O. and {Linz}, H. and {Lippok}, N. and {Ragan}, S. and {Risacher}, C. and {Schmiedeke}, A.},
        title = "{The Earliest Phases of Star formation (EPoS) observed with Herschel: the dust temperature and density distributions of B68}",
      journal = {\aap},
         year = 2012,
        month = nov,
       volume = {547},
          eid = {A11},
        pages = {A11},
          doi = {10.1051/0004-6361/201219139},
archivePrefix = {arXiv},
       eprint = {1208.4512},
 primaryClass = {astro-ph.GA},
       adsurl = {https://ui.adsabs.harvard.edu/abs/2012A&A...547A..11N}
}

@ARTICLE{1992ApJ...393..278L,
       author = {{Lada}, Charles J. and {Adams}, Fred C.},
        title = "{Interpreting Infrared Color-Color Diagrams: Circumstellar Disks around Low- and Intermediate-Mass Young Stellar Objects}",
      journal = {\apj},
         year = 1992,
        month = jul,
       volume = {393},
        pages = {278},
          doi = {10.1086/171505},
       adsurl = {https://ui.adsabs.harvard.edu/abs/1992ApJ...393..278L}
}

@ARTICLE{1983QJRAS..24..267H,
       author = {{Hildebrand}, R.~H.},
        title = "{The determination of cloud masses and dust characteristics from submillimetre thermal emission.}",
      journal = {\qjras},
         year = 1983,
        month = sep,
       volume = {24},
        pages = {267-282},
       adsurl = {https://ui.adsabs.harvard.edu/abs/1983QJRAS..24..267H}
}

@ARTICLE{1997A&A...320L..41H,
       author = {{Horrobin}, M.~J. and {Casali}, M.~M. and {Eiroa}, C.},
        title = "{Variability and a vanishing YSO in the Serpens cloud core.}",
      journal = {\aap},
         year = 1997,
        month = apr,
       volume = {320},
        pages = {L41-L43},
       adsurl = {https://ui.adsabs.harvard.edu/abs/1997A&A...320L..41H}
}

@ARTICLE{2023A&A...674A..32B,
       author = {{Babusiaux}, C. and {Fabricius}, C. and {Khanna}, S. and {Muraveva}, T. and {Reyl{\'e}}, C. and {Spoto}, F. and {Vallenari}, A. and {Luri}, X. and {Arenou}, F. and {{\'A}lvarez}, M.~A. and {Anders}, F. and {Antoja}, T. and {Balbinot}, E. and {Barache}, C. and {Bauchet}, N. and {Bossini}, D. and {Busonero}, D. and {Cantat-Gaudin}, T. and {Carrasco}, J.~M. and {Dafonte}, C. and {Diakit{\'e}}, S. and {Figueras}, F. and {Garcia-Gutierrez}, A. and {Garofalo}, A. and {Helmi}, A. and {Jim{\'e}nez-Arranz}, {\'O}. and {Jordi}, C. and {Kervella}, P. and {Kostrzewa-Rutkowska}, Z. and {Leclerc}, N. and {Licata}, E. and {Manteiga}, M. and {Masip}, A. and {Mongui{\'o}}, M. and {Ramos}, P. and {Robichon}, N. and {Robin}, A.~C. and {Romero-G{\'o}mez}, M. and {S{\'a}ez}, A. and {Santove{\~n}a}, R. and {Spina}, L. and {Torralba Elipe}, G. and {Weiler}, M.},
        title = "{Gaia Data Release 3. Catalogue validation}",
      journal = {\aap},
         year = 2023,
        month = jun,
       volume = {674},
          eid = {A32},
        pages = {A32},
          doi = {10.1051/0004-6361/202243790},
archivePrefix = {arXiv},
       eprint = {2206.05989},
 primaryClass = {astro-ph.SR},
       adsurl = {https://ui.adsabs.harvard.edu/abs/2023A&A...674A..32B}
}

@article{Saral_2015,
doi = {10.1088/0004-637X/813/1/25},
url = {https://dx.doi.org/10.1088/0004-637X/813/1/25},
year = {2015},
month = {oct},
publisher = {The American Astronomical Society},
volume = {813},
number = {1},
pages = {25},
author = {Saral, G. and Hora, J. L. and Willis, S. E. and Koenig, X. P. and Gutermuth, R. A. and Saygac, A. T.},
title = {YOUNG STELLAR OBJECTS IN THE MASSIVE STAR-FORMING REGION W49},
journal = {The Astrophysical Journal}
}

@article{Zhang_2014,
doi = {10.1088/0004-637X/781/2/89},
url = {https://dx.doi.org/10.1088/0004-637X/781/2/89},
year = {2014},
month = {jan},
publisher = {The American Astronomical Society},
volume = {781},
number = {2},
pages = {89},
author = {Zhang, B. and Moscadelli, L. and Sato, M. and Reid, M. J. and Menten, K. M. and Zheng, X. W. and Brunthaler, A. and Dame, T. M. and Xu, Y. and Immer, K.},
title = {THE PARALLAX OF W43: A MASSIVE STAR-FORMING COMPLEX NEAR THE GALACTIC BAR},
journal = {The Astrophysical Journal}
}

@ARTICLE{2011A&A...529A..41N,
       author = {{Nguyen Luong}, Q. and {Motte}, F. and {Schuller}, F. and {Schneider}, N. and {Bontemps}, S. and {Schilke}, P. and {Menten}, K.~M. and {Heitsch}, F. and {Wyrowski}, F. and {Carlhoff}, P. and {Bronfman}, L. and {Henning}, T.},
        title = "{W43: the closest molecular complex of the Galactic bar?}",
      journal = {\aap},
         year = 2011,
        month = may,
       volume = {529},
          eid = {A41},
        pages = {A41},
          doi = {10.1051/0004-6361/201016271},
archivePrefix = {arXiv},
       eprint = {1102.3460},
 primaryClass = {astro-ph.SR},
       adsurl = {https://ui.adsabs.harvard.edu/abs/2011A&A...529A..41N}
}

@ARTICLE{2021ApJS..254...33K,
       author = {{Kuhn}, Michael A. and {de Souza}, Rafael S. and {Krone-Martins}, Alberto and {Castro-Ginard}, Alfred and {Ishida}, Emille E.~O. and {Povich}, Matthew S. and {Hillenbrand}, Lynne A. and {COIN Collaboration}},
        title = "{SPICY: The Spitzer/IRAC Candidate YSO Catalog for the Inner Galactic Midplane}",
      journal = {\apjs},
         year = 2021,
        month = jun,
       volume = {254},
       number = {2},
          eid = {33},
        pages = {33},
          doi = {10.3847/1538-4365/abe465},
archivePrefix = {arXiv},
       eprint = {2011.12961},
 primaryClass = {astro-ph.GA},
       adsurl = {https://ui.adsabs.harvard.edu/abs/2021ApJS..254...33K}
}

@ARTICLE{2006MNRAS.371..703S,
       author = {{S{\'a}nchez-Bl{\'a}zquez}, P. and {Peletier}, R.~F. and {Jim{\'e}nez-Vicente}, J. and {Cardiel}, N. and {Cenarro}, A.~J. and {Falc{\'o}n-Barroso}, J. and {Gorgas}, J. and {Selam}, S. and {Vazdekis}, A.},
        title = "{Medium-resolution Isaac Newton Telescope library of empirical spectra}",
      journal = {\mnras},
         year = 2006,
        month = sep,
       volume = {371},
       number = {2},
        pages = {703-718},
          doi = {10.1111/j.1365-2966.2006.10699.x},
archivePrefix = {arXiv},
       eprint = {astro-ph/0607009},
 primaryClass = {astro-ph},
       adsurl = {https://ui.adsabs.harvard.edu/abs/2006MNRAS.371..703S}
}

@ARTICLE{2016A&A...595A...2G,
       author = {{Gaia Collaboration} and {Brown}, A.~G.~A. and {Vallenari}, A. and {Prusti}, T. and {de Bruijne}, J.~H.~J. and {Mignard}, F. and {Drimmel}, R. and {Babusiaux}, C. and {Bailer-Jones}, C.~A.~L. and {Bastian}, U. and {Biermann}, M. and {Evans}, D.~W. and {Eyer}, L. and {Jansen}, F. and {Jordi}, C. and {Katz}, D. and {Klioner}, S.~A. and {Lammers}, U. and {Lindegren}, L. and {Luri}, X. and {O'Mullane}, W. and {Panem}, C. and {Pourbaix}, D. and {Randich}, S. and {Sartoretti}, P. and {Siddiqui}, H.~I. and {Soubiran}, C. and {Valette}, V. and {van Leeuwen}, F. and {Walton}, N.~A. and {Aerts}, C. and {Arenou}, F. and {Cropper}, M. and {H{\o}g}, E. and {Lattanzi}, M.~G. and {Grebel}, E.~K. and {Holland}, A.~D. and {Huc}, C. and {Passot}, X. and {Perryman}, M. and {Bramante}, L. and {Cacciari}, C. and {Casta{\~n}eda}, J. and {Chaoul}, L. and {Cheek}, N. and {De Angeli}, F. and {Fabricius}, C. and {Guerra}, R. and {Hern{\'a}ndez}, J. and {Jean-Antoine-Piccolo}, A. and {Masana}, E. and {Messineo}, R. and {Mowlavi}, N. and {Nienartowicz}, K. and {Ord{\'o}{\~n}ez-Blanco}, D. and {Panuzzo}, P. and {Portell}, J. and {Richards}, P.~J. and {Riello}, M. and {Seabroke}, G.~M. and {Tanga}, P. and {Th{\'e}venin}, F. and {Torra}, J. and {Els}, S.~G. and {Gracia-Abril}, G. and {Comoretto}, G. and {Garcia-Reinaldos}, M. and {Lock}, T. and {Mercier}, E. and {Altmann}, M. and {Andrae}, R. and {Astraatmadja}, T.~L. and {Bellas-Velidis}, I. and {Benson}, K. and {Berthier}, J. and {Blomme}, R. and {Busso}, G. and {Carry}, B. and {Cellino}, A. and {Clementini}, G. and {Cowell}, S. and {Creevey}, O. and {Cuypers}, J. and {Davidson}, M. and {De Ridder}, J. and {de Torres}, A. and {Delchambre}, L. and {Dell'Oro}, A. and {Ducourant}, C. and {Fr{\'e}mat}, Y. and {Garc{\'\i}a-Torres}, M. and {Gosset}, E. and {Halbwachs}, J. -L. and {Hambly}, N.~C. and {Harrison}, D.~L. and {Hauser}, M. and {Hestroffer}, D. and {Hodgkin}, S.~T. and {Huckle}, H.~E. and {Hutton}, A. and {Jasniewicz}, G. and {Jordan}, S. and {Kontizas}, M. and {Korn}, A.~J. and {Lanzafame}, A.~C. and {Manteiga}, M. and {Moitinho}, A. and {Muinonen}, K. and {Osinde}, J. and {Pancino}, E. and {Pauwels}, T. and {Petit}, J. -M. and {Recio-Blanco}, A. and {Robin}, A.~C. and {Sarro}, L.~M. and {Siopis}, C. and {Smith}, M. and {Smith}, K.~W. and {Sozzetti}, A. and {Thuillot}, W. and {van Reeven}, W. and {Viala}, Y. and {Abbas}, U. and {Abreu Aramburu}, A. and {Accart}, S. and {Aguado}, J.~J. and {Allan}, P.~M. and {Allasia}, W. and {Altavilla}, G. and {{\'A}lvarez}, M.~A. and {Alves}, J. and {Anderson}, R.~I. and {Andrei}, A.~H. and {Anglada Varela}, E. and {Antiche}, E. and {Antoja}, T. and {Ant{\'o}n}, S. and {Arcay}, B. and {Bach}, N. and {Baker}, S.~G. and {Balaguer-N{\'u}{\~n}ez}, L. and {Barache}, C. and {Barata}, C. and {Barbier}, A. and {Barblan}, F. and {Barrado y Navascu{\'e}s}, D. and {Barros}, M. and {Barstow}, M.~A. and {Becciani}, U. and {Bellazzini}, M. and {Bello Garc{\'\i}a}, A. and {Belokurov}, V. and {Bendjoya}, P. and {Berihuete}, A. and {Bianchi}, L. and {Bienaym{\'e}}, O. and {Billebaud}, F. and {Blagorodnova}, N. and {Blanco-Cuaresma}, S. and {Boch}, T. and {Bombrun}, A. and {Borrachero}, R. and {Bouquillon}, S. and {Bourda}, G. and {Bouy}, H. and {Bragaglia}, A. and {Breddels}, M.~A. and {Brouillet}, N. and {Br{\"u}semeister}, T. and {Bucciarelli}, B. and {Burgess}, P. and {Burgon}, R. and {Burlacu}, A. and {Busonero}, D. and {Buzzi}, R. and {Caffau}, E. and {Cambras}, J. and {Campbell}, H. and {Cancelliere}, R. and {Cantat-Gaudin}, T. and {Carlucci}, T. and {Carrasco}, J.~M. and {Castellani}, M. and {Charlot}, P. and {Charnas}, J. and {Chiavassa}, A. and {Clotet}, M. and {Cocozza}, G. and {Collins}, R.~S. and {Costigan}, G. and {Crifo}, F. and {Cross}, N.~J.~G. and {Crosta}, M. and {Crowley}, C. and {Dafonte}, C. and {Damerdji}, Y. and {Dapergolas}, A. and {David}, P. and {David}, M. and {De Cat}, P.},
        title = "{Gaia Data Release 1. Summary of the astrometric, photometric, and survey properties}",
      journal = {\aap},
         year = 2016,
        month = nov,
       volume = {595},
          eid = {A2},
        pages = {A2},
          doi = {10.1051/0004-6361/201629512},
archivePrefix = {arXiv},
       eprint = {1609.04172},
 primaryClass = {astro-ph.IM},
       adsurl = {https://ui.adsabs.harvard.edu/abs/2016A&A...595A...2G}
}

@ARTICLE{2003PASP..115..953B,
       author = {{Benjamin}, Robert A. and {Churchwell}, E. and {Babler}, Brian L. and {Bania}, T.~M. and {Clemens}, Dan P. and {Cohen}, Martin and {Dickey}, John M. and {Indebetouw}, R{\'e}my and {Jackson}, James M. and {Kobulnicky}, Henry A. and {Lazarian}, Alex and {Marston}, A.~P. and {Mathis}, John S. and {Meade}, Marilyn R. and {Seager}, Sara and {Stolovy}, S.~R. and {Watson}, C. and {Whitney}, Barbara A. and {Wolff}, Michael J. and {Wolfire}, Mark G.},
        title = "{GLIMPSE. I. An SIRTF Legacy Project to Map the Inner Galaxy}",
      journal = {\pasp},
         year = 2003,
        month = aug,
       volume = {115},
       number = {810},
        pages = {953-964},
          doi = {10.1086/376696},
archivePrefix = {arXiv},
       eprint = {astro-ph/0306274},
 primaryClass = {astro-ph},
       adsurl = {https://ui.adsabs.harvard.edu/abs/2003PASP..115..953B}
}

@ARTICLE{2009PASP..121..213C,
       author = {{Churchwell}, Ed and {Babler}, Brian L. and {Meade}, Marilyn R. and {Whitney}, Barbara A. and {Benjamin}, Robert and {Indebetouw}, Remy and {Cyganowski}, Claudia and {Robitaille}, Thomas P. and {Povich}, Matthew and {Watson}, Christer and {Bracker}, Steve},
        title = "{The Spitzer/GLIMPSE Surveys: A New View of the Milky Way}",
      journal = {\pasp},
         year = 2009,
        month = mar,
       volume = {121},
       number = {877},
        pages = {213},
          doi = {10.1086/597811},
       adsurl = {https://ui.adsabs.harvard.edu/abs/2009PASP..121..213C}
}

@ARTICLE{2010A&A...518L...1P,
       author = {{Pilbratt}, G.~L. and {Riedinger}, J.~R. and {Passvogel}, T. and {Crone}, G. and {Doyle}, D. and {Gageur}, U. and {Heras}, A.~M. and {Jewell}, C. and {Metcalfe}, L. and {Ott}, S. and {Schmidt}, M.},
        title = "{Herschel Space Observatory. An ESA facility for far-infrared and submillimetre astronomy}",
      journal = {\aap},
         year = 2010,
        month = jul,
       volume = {518},
          eid = {L1},
        pages = {L1},
          doi = {10.1051/0004-6361/201014759},
archivePrefix = {arXiv},
       eprint = {1005.5331},
 primaryClass = {astro-ph.IM},
       adsurl = {https://ui.adsabs.harvard.edu/abs/2010A&A...518L...1P}
}

@ARTICLE{2010A&A...518L...2P,
       author = {{Poglitsch}, A. and {Waelkens}, C. and {Geis}, N. and {Feuchtgruber}, H. and {Vandenbussche}, B. and {Rodriguez}, L. and {Krause}, O. and {Renotte}, E. and {van Hoof}, C. and {Saraceno}, P. and {Cepa}, J. and {Kerschbaum}, F. and {Agn{\`e}se}, P. and {Ali}, B. and {Altieri}, B. and {Andreani}, P. and {Augueres}, J. -L. and {Balog}, Z. and {Barl}, L. and {Bauer}, O.~H. and {Belbachir}, N. and {Benedettini}, M. and {Billot}, N. and {Boulade}, O. and {Bischof}, H. and {Blommaert}, J. and {Callut}, E. and {Cara}, C. and {Cerulli}, R. and {Cesarsky}, D. and {Contursi}, A. and {Creten}, Y. and {De Meester}, W. and {Doublier}, V. and {Doumayrou}, E. and {Duband}, L. and {Exter}, K. and {Genzel}, R. and {Gillis}, J. -M. and {Gr{\"o}zinger}, U. and {Henning}, T. and {Herreros}, J. and {Huygen}, R. and {Inguscio}, M. and {Jakob}, G. and {Jamar}, C. and {Jean}, C. and {de Jong}, J. and {Katterloher}, R. and {Kiss}, C. and {Klaas}, U. and {Lemke}, D. and {Lutz}, D. and {Madden}, S. and {Marquet}, B. and {Martignac}, J. and {Mazy}, A. and {Merken}, P. and {Montfort}, F. and {Morbidelli}, L. and {M{\"u}ller}, T. and {Nielbock}, M. and {Okumura}, K. and {Orfei}, R. and {Ottensamer}, R. and {Pezzuto}, S. and {Popesso}, P. and {Putzeys}, J. and {Regibo}, S. and {Reveret}, V. and {Royer}, P. and {Sauvage}, M. and {Schreiber}, J. and {Stegmaier}, J. and {Schmitt}, D. and {Schubert}, J. and {Sturm}, E. and {Thiel}, M. and {Tofani}, G. and {Vavrek}, R. and {Wetzstein}, M. and {Wieprecht}, E. and {Wiezorrek}, E.},
        title = "{The Photodetector Array Camera and Spectrometer (PACS) on the Herschel Space Observatory}",
      journal = {\aap},
         year = 2010,
        month = jul,
       volume = {518},
          eid = {L2},
        pages = {L2},
          doi = {10.1051/0004-6361/201014535},
archivePrefix = {arXiv},
       eprint = {1005.1487},
 primaryClass = {astro-ph.IM},
       adsurl = {https://ui.adsabs.harvard.edu/abs/2010A&A...518L...2P}
}

@ARTICLE{2010A&A...518L...3G,
       author = {{Griffin}, M.~J. and {Abergel}, A. and {Abreu}, A. and {Ade}, P.~A.~R. and {Andr{\'e}}, P. and {Augueres}, J. -L. and {Babbedge}, T. and {Bae}, Y. and {Baillie}, T. and {Baluteau}, J. -P. and {Barlow}, M.~J. and {Bendo}, G. and {Benielli}, D. and {Bock}, J.~J. and {Bonhomme}, P. and {Brisbin}, D. and {Brockley-Blatt}, C. and {Caldwell}, M. and {Cara}, C. and {Castro-Rodriguez}, N. and {Cerulli}, R. and {Chanial}, P. and {Chen}, S. and {Clark}, E. and {Clements}, D.~L. and {Clerc}, L. and {Coker}, J. and {Communal}, D. and {Conversi}, L. and {Cox}, P. and {Crumb}, D. and {Cunningham}, C. and {Daly}, F. and {Davis}, G.~R. and {de Antoni}, P. and {Delderfield}, J. and {Devin}, N. and {di Giorgio}, A. and {Didschuns}, I. and {Dohlen}, K. and {Donati}, M. and {Dowell}, A. and {Dowell}, C.~D. and {Duband}, L. and {Dumaye}, L. and {Emery}, R.~J. and {Ferlet}, M. and {Ferrand}, D. and {Fontignie}, J. and {Fox}, M. and {Franceschini}, A. and {Frerking}, M. and {Fulton}, T. and {Garcia}, J. and {Gastaud}, R. and {Gear}, W.~K. and {Glenn}, J. and {Goizel}, A. and {Griffin}, D.~K. and {Grundy}, T. and {Guest}, S. and {Guillemet}, L. and {Hargrave}, P.~C. and {Harwit}, M. and {Hastings}, P. and {Hatziminaoglou}, E. and {Herman}, M. and {Hinde}, B. and {Hristov}, V. and {Huang}, M. and {Imhof}, P. and {Isaak}, K.~J. and {Israelsson}, U. and {Ivison}, R.~J. and {Jennings}, D. and {Kiernan}, B. and {King}, K.~J. and {Lange}, A.~E. and {Latter}, W. and {Laurent}, G. and {Laurent}, P. and {Leeks}, S.~J. and {Lellouch}, E. and {Levenson}, L. and {Li}, B. and {Li}, J. and {Lilienthal}, J. and {Lim}, T. and {Liu}, S.~J. and {Lu}, N. and {Madden}, S. and {Mainetti}, G. and {Marliani}, P. and {McKay}, D. and {Mercier}, K. and {Molinari}, S. and {Morris}, H. and {Moseley}, H. and {Mulder}, J. and {Mur}, M. and {Naylor}, D.~A. and {Nguyen}, H. and {O'Halloran}, B. and {Oliver}, S. and {Olofsson}, G. and {Olofsson}, H. -G. and {Orfei}, R. and {Page}, M.~J. and {Pain}, I. and {Panuzzo}, P. and {Papageorgiou}, A. and {Parks}, G. and {Parr-Burman}, P. and {Pearce}, A. and {Pearson}, C. and {P{\'e}rez-Fournon}, I. and {Pinsard}, F. and {Pisano}, G. and {Podosek}, J. and {Pohlen}, M. and {Polehampton}, E.~T. and {Pouliquen}, D. and {Rigopoulou}, D. and {Rizzo}, D. and {Roseboom}, I.~G. and {Roussel}, H. and {Rowan-Robinson}, M. and {Rownd}, B. and {Saraceno}, P. and {Sauvage}, M. and {Savage}, R. and {Savini}, G. and {Sawyer}, E. and {Scharmberg}, C. and {Schmitt}, D. and {Schneider}, N. and {Schulz}, B. and {Schwartz}, A. and {Shafer}, R. and {Shupe}, D.~L. and {Sibthorpe}, B. and {Sidher}, S. and {Smith}, A. and {Smith}, A.~J. and {Smith}, D. and {Spencer}, L. and {Stobie}, B. and {Sudiwala}, R. and {Sukhatme}, K. and {Surace}, C. and {Stevens}, J.~A. and {Swinyard}, B.~M. and {Trichas}, M. and {Tourette}, T. and {Triou}, H. and {Tseng}, S. and {Tucker}, C. and {Turner}, A. and {Vaccari}, M. and {Valtchanov}, I. and {Vigroux}, L. and {Virique}, E. and {Voellmer}, G. and {Walker}, H. and {Ward}, R. and {Waskett}, T. and {Weilert}, M. and {Wesson}, R. and {White}, G.~J. and {Whitehouse}, N. and {Wilson}, C.~D. and {Winter}, B. and {Woodcraft}, A.~L. and {Wright}, G.~S. and {Xu}, C.~K. and {Zavagno}, A. and {Zemcov}, M. and {Zhang}, L. and {Zonca}, E.},
        title = "{The Herschel-SPIRE instrument and its in-flight performance}",
      journal = {\aap},
         year = 2010,
        month = jul,
       volume = {518},
          eid = {L3},
        pages = {L3},
          doi = {10.1051/0004-6361/201014519},
archivePrefix = {arXiv},
       eprint = {1005.5123},
 primaryClass = {astro-ph.IM},
       adsurl = {https://ui.adsabs.harvard.edu/abs/2010A&A...518L...3G}
}

@ARTICLE{1998A&AS..133..387B,
       author = {{Balaguer-N{\'u}nez}, L. and {Tian}, K.~P. and {Zhao}, J.~L.},
        title = "{Determination of proper motions and membership of the open clusters NGC 1817 and NGC 1807}",
      journal = {\aaps},
         year = 1998,
        month = dec,
       volume = {133},
        pages = {387-394},
          doi = {10.1051/aas:1998324},
       adsurl = {https://ui.adsabs.harvard.edu/abs/1998A&AS..133..387B}
}

@ARTICLE{1989AJ.....98..227G,
       author = {{Girard}, Terrence M. and {Grundy}, William M. and {Lopez}, Carlos E. and {van Altena}, William F.},
        title = "{Relative Proper Motions and the Stellar Velocity Dispersion of the Open Cluster M67}",
      journal = {\aj},
         year = 1989,
        month = jul,
       volume = {98},
        pages = {227},
          doi = {10.1086/115139},
       adsurl = {https://ui.adsabs.harvard.edu/abs/1989AJ.....98..227G}
}

@article{Bisht_2022,
doi = {10.1088/1538-3873/ac6195},
url = {https://dx.doi.org/10.1088/1538-3873/ac6195},
year = {2022},
month = {apr},
publisher = {The Astronomical Society of the Pacific},
volume = {134},
number = {1034},
pages = {044201},
author = {Bisht, D. and Zhu, Qingfeng and Yadav, R. K. S. and Rangwal, Geeta and Sariya, Devesh P. and Durgapal, Alok and Jiang, Ing-Guey},
title = {A Deep Investigation of Two Poorly Studied Open Clusters Haffner 22 and Melotte 71 in the Gaia era},
journal = {Publications of the Astronomical Society of the Pacific}
}

@article{10.1093/mnras/stt136,
    author = {Yadav, R. K. S. and Sariya, Devesh P. and Sagar, R.},
    title = {Proper motions and membership probabilities of stars in the region of open cluster NGC 3766},
    journal = {Monthly Notices of the Royal Astronomical Society},
    volume = {430},
    number = {4},
    pages = {3350-3358},
    year = {2013},
    month = {02},
    issn = {0035-8711},
    doi = {10.1093/mnras/stt136},
    url = {https://doi.org/10.1093/mnras/stt136},
    eprint = {https://academic.oup.com/mnras/article-pdf/430/4/3350/3898323/stt136.pdf},
}

@BOOK{2000asqu.book.....C,
       author = {{Cox}, Arthur N.},
        title = "{Allen's astrophysical quantities}",
         year = 2000,
       adsurl = {https://ui.adsabs.harvard.edu/abs/2000asqu.book.....C}
}

@ARTICLE{1990AJ.....99..924B,
       author = {{Beckwith}, Steven V.~W. and {Sargent}, Anneila I. and {Chini}, Rolf S. and {Guesten}, Rolf},
        title = "{A Survey for Circumstellar Disks around Young Stellar Objects}",
      journal = {\aj},
         year = 1990,
        month = mar,
       volume = {99},
        pages = {924},
          doi = {10.1086/115385},
       adsurl = {https://ui.adsabs.harvard.edu/abs/1990AJ.....99..924B}
}

@ARTICLE{2010A&A...518L.102A,
       author = {{Andr{\'e}}, Ph. and {Men'shchikov}, A. and {Bontemps}, S. and {K{\"o}nyves}, V. and {Motte}, F. and {Schneider}, N. and {Didelon}, P. and {Minier}, V. and {Saraceno}, P. and {Ward-Thompson}, D. and {di Francesco}, J. and {White}, G. and {Molinari}, S. and {Testi}, L. and {Abergel}, A. and {Griffin}, M. and {Henning}, Th. and {Royer}, P. and {Mer{\'\i}n}, B. and {Vavrek}, R. and {Attard}, M. and {Arzoumanian}, D. and {Wilson}, C.~D. and {Ade}, P. and {Aussel}, H. and {Baluteau}, J. -P. and {Benedettini}, M. and {Bernard}, J. -Ph. and {Blommaert}, J.~A.~D.~L. and {Cambr{\'e}sy}, L. and {Cox}, P. and {di Giorgio}, A. and {Hargrave}, P. and {Hennemann}, M. and {Huang}, M. and {Kirk}, J. and {Krause}, O. and {Launhardt}, R. and {Leeks}, S. and {Le Pennec}, J. and {Li}, J.~Z. and {Martin}, P.~G. and {Maury}, A. and {Olofsson}, G. and {Omont}, A. and {Peretto}, N. and {Pezzuto}, S. and {Prusti}, T. and {Roussel}, H. and {Russeil}, D. and {Sauvage}, M. and {Sibthorpe}, B. and {Sicilia-Aguilar}, A. and {Spinoglio}, L. and {Waelkens}, C. and {Woodcraft}, A. and {Zavagno}, A.},
        title = "{From filamentary clouds to prestellar cores to the stellar IMF: Initial highlights from the Herschel Gould Belt Survey}",
      journal = {\aap},
         year = 2010,
        month = jul,
       volume = {518},
          eid = {L102},
        pages = {L102},
          doi = {10.1051/0004-6361/201014666},
archivePrefix = {arXiv},
       eprint = {1005.2618},
 primaryClass = {astro-ph.GA},
       adsurl = {https://ui.adsabs.harvard.edu/abs/2010A&A...518L.102A}
}
\end{document}